\documentclass[11pt]{article}

\usepackage{fullpage}               
\usepackage{graphicx,epsf,subfigure}
\usepackage{pstricks,pst-node,psfrag}
\usepackage{algorithm,algorithmic}
\usepackage{setspace}
\usepackage{amsfonts}
\usepackage{amsthm, amsmath, bbm, amssymb}
\usepackage{dsfont}
\usepackage{multirow}
\usepackage{bm,url}
\usepackage{threeparttable,booktabs}
\usepackage{verbatim,url}
\usepackage{fancyhdr}
\usepackage{natbib}
\RequirePackage[OT1]{fontenc}
\RequirePackage[colorlinks,citecolor=blue,urlcolor=blue]{hyperref}

\usepackage{todonotes}
\usepackage{threeparttable,booktabs}

\newcommand\Tstrut{\rule{0pt}{2.6ex}}         
\newcommand\Bstrut{\rule[-0.9ex]{0pt}{0pt}}   
\renewcommand{\Pr}[0]{\mathbb{P}}
\newcommand{\phylogeny}[0]{\mathcal{F}}
\newcommand{\Allparameter}[0]{\boldsymbol{\phi}}
\newcommand{\nTips}[0]{N}

\newcommand{\momentum}[0]{\mathbf{p}}

\newcommand{\below}[1]{\mathbf{Y}_{\lfloor #1 \rfloor}}
\newcommand{\abbove}[1]{\mathbf{Y}_{\lceil #1 \rceil}}
\newcommand{\parent}[1]{\mbox{pa}{(#1)}}
\newcommand{\cDensity}[2]{\ensuremath{\Pr(#1 \,|\,#2)}}
\newcommand{\jDensity}[2]{\ensuremath{\Pr(#1 , #2)}}
\newcommand{\p}[1]{\mathbf{p}_{#1}}
\newcommand{\q}[1]{\mathbf{q}_{#1}}
\newcommand{\Ptr}[1]{\mathbf{P}_{#1}}
\newcommand{\gradient}[1]{\frac{\partial}{\partial #1}}

\newcommand{\lnP}[1]{\ensuremath{\log \Pr ({#1})}}
\newcommand{\bl}[1]{{b}_{#1}} 
\newcommand{\br}[1]{{r}_{#1}} 
\newcommand{\sr}[1]{{\gamma}_{#1}} 
\newcommand{\nodeT}[1]{t_{#1}} 
\newcommand{\mass}{\mathbf{M}}
\newcommand{\transpose}{^{\prime}}
\newcommand{\dQ}[1]{\partial\mathbf{Q}_{#1}/\partial\theta_{#1}}
\newcommand{\dQv}[1]{\frac{\partial\mathbf{Q}_{#1}}{\partial\theta_{#1}}}

\renewcommand{\circ}{\odot}

\newwrite\XTR
\AtBeginDocument{\immediate\openout\XTR\jobname.xtr}
\AtEndDocument{\immediate\closeout\XTR}

\newcommand{\defineXtrCommand}[2]{%
	\write\XTR{%
		\string\newcommand%
		{\csname myedit#1\endcsname}%
	}%
	\write\XTR{{#2}}%
}

\newcommand{\myedit}[2]{
	\defineXtrCommand{#1}{%
		``\expandafter\string#2''
		\null \hfill \mbox{(Page\string~\thepage)}%
	}%
	\defineXtrCommand{#1String}{%
		\expandafter\string#2%
	}%
	\defineXtrCommand{#1Page}{%
		\thepage%
	}%
	#2%
}

\newcommand{\myeditNoPrint}[2]{
	\defineXtrCommand{#1}{%
		``\expandafter\detokenize{\string#2}''
		\null \hfill (Page\string~\thepage)%
	}%
	\defineXtrCommand{#1String}{%
		\expandafter\string#2%
	}%
	\defineXtrCommand{#1Page}{%
		\thepage%
	}%
}

\begin{document}
	
	\begin{flushright}
		Article (Methods)\\
	\end{flushright}

	\begin{center}
		\begin{Large}
			{\bf Detecting Evolutionary Change-Points with Branch-Specific Substitution Models and Shrinkage Priors}
		\end{Large}
	\end{center}
	

\begin{center}
	{Xiang Ji$^{\ast, 1}$, Benjamin Redelings$^{2}$, Shuo Su$^{3}$, Hongcun Bao$^{4}$, Wu-Min Deng$^{4}$, Filippo Monti$^{5}$, Samuel L. Hong$^{6}$, Guy Baele$^{6}$, Philippe Lemey$^{6}$, and Marc A. Suchard$^{\ast, 5, 7, 8}$\\

		\small $^{1}$Department of Statistics, College of Liberal Arts and Sciences,\\
		Iowa State University, Ames, IA, USA\\
		\small $^{2}$Department of Mathematics, School of Science and Engineering,\\
		Tulane University, New Orleans, LA, USA\\
		\small $^{3}$Shanghai Institute of Infectious Disease and Biosecurity, School of Public Health,\\
		Fudan University, Shanghai, China\\
		\small $^{4}$Department of Biochemistry and Molecular Biology,\\
		Tulane University, New Orleans, LA, USA\\
		\small $^{5}$Department of Biostatistics, Fielding School of Public Health,\\
		University of California Los Angeles, Los Angeles, CA, USA\\
		\small $^{6}$Department of Microbiology, Immunology and Transplantation, Rega Institute,\\
		KU Leuven, Leuven, Belgium\\
		\small $^{7}$Department of Biomathematics and
		\small $^{8}$Department of Human Genetics,\\
		David Geffen School of Medicine,
		University of California Los Angeles, Los Angeles, CA, USA\\
		\small $^{*}$Correspondence: xiangji@iastate.edu, msuchard@ucla.edu
}\end{center}

	\date{}
	
	
	\begin{abstract}
		Branch-specific substitution models are popular for detecting evolutionary change-points, such as shifts in selective pressure.
		However, applying such models typically requires prior knowledge of change-point locations on the phylogeny or faces scalability issues with large data sets.
		To address both limitations, we integrate branch-specific substitution models with shrinkage priors to automatically identify change-points without prior knowledge, while simultaneously estimating distinct substitution parameters for each branch.
		To enable tractable inference under this high-dimensional model, we develop an analytical gradient algorithm for the branch-specific substitution parameters where the computational time is linear in the number of parameters.
		We apply this gradient algorithm to infer selection pressure dynamics in the evolution of the BRCA1 gene in primates and mutational dynamics in viral sequences from the recent mpox epidemic.
		Our novel algorithm enhances inference efficiency, achieving up to a 126-fold speedup per iteration in maximum likelihood optimization when compared to central difference numerical gradient method and up to a 2026-fold improvement in computational performance within a Bayesian framework using Hamiltonian Monte Carlo sampler compared to conventional univariate random walk sampler.
	\end{abstract}
	
	
	\section{Introduction}
	A striking diversity exists in size, life history, ecology, population structure, physiology, and cellular biology among major groups of organisms.
	This diversity extends to the genomic level, where substantial interspecies variation in sequence substitution rates and base frequencies is observed.
	However, standard evolutionary models often assume a homogeneous substitution process across all branches of a phylogenetic tree \citep{muse1994likelihood, goldman1994codon}. 
	This assumption is frequently violated in real-world scenarios, particularly in rapidly evolving pathogens like viruses, where evolutionary pressures can vary substantially across lineages due to host shifts, immune escape, drug resistance, or environmental changes \citep{holmes2009evolution, lemey2006hiv, wertheim2014global}. 
	The evolution of functional protein-coding genes constitutes another example where potential shifts in evolutionary dynamics could be associated with macroscopic trait changes.
	Nonsynonymous changes to protein-coding DNA affect the resulting amino acid sequence, whereas synonymous changes do not.
	Since natural selection mainly operates at the protein level, the $\omega$ parameter in the GY94 codon substitution model \citep{goldman1994codon} that represents the ratio of nonsynonymous to synonymous codon substitution rates ($dN/dS$) can serve as an informative measure of natural selection.
	Statistical models that incorporate branch-specific substitution processes offer a flexible and robust framework for detecting such evolutionary changes \citep{yang2002codon, huelsenbeck2000compound, pond2005genetic}. 

	Despite their biological relevance, branch-specific models pose substantial statistical challenges, including increased model complexity, risk of overfitting, and the need for efficient computational strategies \citep{baele2012improving, baele2016bayesian, hohna2019bayesian}. 
	Many studies of lineage-specific variation in selective pressure rely on PAML \citep{yang2007paml} and use one of two distinct procedures: (1) the user specifies \textit{a priori} clusters of branches on a phylogeny to share a similar $dN/dS$ or (2) assumes that each branch has its own $dN/dS$.
	The first option requires prior knowledge of change-point locations on the tree to group branches to share the same or similar $dN/dS$ values.
	The second option on the other hand  is discouraged by the PAML manual as it estimates one parameter for every branch of the tree and tends to be unstable when the tree is large.
\myedit{introLinearComplexity}{%
		To address these challenges and enable tractable inference under high-dimensional models, we develop a linear (in the number of tips) computational time  analytical gradient algorithm that efficiently computes the derivatives of the log likelihood function with respect to (w.r.t.) all branch-specific substitution parameters at once.%
}
	This algorithm enables the application of recent advances in Bayesian phylogenetic inference, such as Hamiltonian Monte Carlo (HMC) sampling and shrinkage priors, to facilitate more robust estimation of branch-wise heterogeneity and improve the reliability of these models \citep{ji2020gradients, fisher2021relaxed, fisher2023shrinkage}.
	We demonstrate these improvements to infer the selection pressure dynamics in the evolution of the interspecies protein-coding BRCA1 gene in primates \citep{yang2002codon} and the mutational dynamics in the evolution of mpox virus sequences from a recent outbreak \citep{o2023apobec3}.


        
	\section{New Approach} \label{sec:algorithm}
	In this section, we derive a new analytic gradient algorithm w.r.t.~substitution parameters.
	This new exact gradient algorithm coupled with 
	shrinkage priors and HMC enables scalable Bayesian phylogenetic inference with branch-specific substitution processes without a need to fix the tree topology or prior knowledge of change-points on the tree.
	The new gradient algorithm leverages the  post- and pre-order partial likelihood vectors from \cite{ji2020gradients} that we briefly review, as well as the numeric implementation in the high-performance  BEAGLE library \citep{ayres2019beagle, gangavarapu2024many} for calculating the likelihood and its gradients.

We start by introducing necessary notations and a brief review of the likelihood calculation with post-order traversals.
We then derive the $O(\nTips)$-dimensional gradient algorithm for substitution parameters (other than the branch lengths) using the post- and pre-order partial likelihood vectors and differential of the instantaneous transition matrices.
We demonstrate two specific forms of the gradient w.r.t.~substitution parameters of a 4-state nucleotide substitution model and a 61-state codon substitution model.
Finally, we review the autocorrelated shrinkage Bayesian bridge prior and HMC methods previously employed to learn evolutionary rates in \cite{fisher2023shrinkage} to complete the section.

	
	\subsection{Notation} \label{sec:notations}
	We follow the notation defined in \cite{ji2020gradients} and derive the gradient algorithm.
	Let $\phylogeny$ represent a phylogeny with $\nTips$ tips and $\nTips-1$ internal nodes.
	We place the root node on the top and the tip nodes at the bottom of $\phylogeny$.
	We denote each node with a number such that $1, 2, ..., \nTips$ are for tip nodes and $\nTips+1, \nTips+2, ..., 2\nTips-1$ are for internal nodes where the root node is fixed at $2\nTips-1$.
        We denote the parent of node $i$ as $\parent{i}$.
        We refer to the branch $(i, \parent{i})$ by the number $i$ of the child node.
	On $\phylogeny$, we model the columns (i.e.,~sites) in the sequence alignment as independent and identically distributed such that they arise from conditionally independent continuous-time Markov chains (CTMCs)  acting along each branch.
	More specifically, throughout this manuscript, we consider non-homogeneous CTMCs acting on $\phylogeny$ where each branch has its own and possibly different CTMC processes.
	
	We denote the state on node $i$ at a site (i.e.,~single column of the sequence alignment) by $Y_i$.
	To ease the presentation, all our derivation will be for a single site and will omit the summation over sites.
	Our derivation naturally generalizes to multiple sites with across-site-variation via discretized rate categories \citep[see e.g.,~Section~\ref{sec:rate_heterogenity};][]{yang1994maximum}  or Markov-modulated modeling for competing Markov processes \citep{baele2021markov}.
	We 
	consider a state space of size $m$ (e.g.,~$m=4$ for nucleotide substitution models, $m=20$ for amino acid substitution models and $m=61$ for codon substitution models that exclude the stop codons).
	We denote the branch length and the evolutionary rate of branch $i$ by $\bl{i}$ and $\br{i}$ respectively, and the real time of node $i$ by $\nodeT{i}$,
	such that $\bl{i} = \br{i} (\nodeT{i} - \nodeT{\parent{i}})$.
	For branch $i$, we denote its infinitesimal rate matrix by $\mathbf{Q}_i \in \mathbb{R}^{m \times m}$ 
	and its eigen decomposition by $\mathbf{Q}_i = \mathbf{U}_i\boldsymbol{\Lambda}_i\mathbf{U}_i^{-1}$ such that $\mathbf{U}_i \in \mathbb{R}^{m \times m}$ whose columns are the eigenvectors of $\mathbf{Q}_i$ and $\boldsymbol{\Lambda}_i = diag\{\lambda_{i1}, \lambda_{i2}, \dots, \lambda_{im}\}$ is a diagonal matrix containing real eigenvalues (see~Supplementary Material for detailed derivation that includes complex eigenvalues).
	The transition probability matrix $\Ptr{i}$ of branch $i$ becomes  $\Ptr{i} = e^{\mathbf{Q}_i \bl{i}} = \mathbf{U}_i e^{\boldsymbol{\Lambda}_i \bl{i}} \mathbf{U}_i^{-1}$.
	We denote the state distribution at the root node by $\boldsymbol{\pi} = \left[ \Pr(Y_{2\nTips-1} = 1), \Pr(Y_{2\nTips-1} = 2), \ldots, \Pr(Y_{2\nTips-1} = m) \right]\transpose$ %
		\myedit{firstTranspose}{%
		(not necessarily the stationary distribution of the CTMCs) where $\transpose$ indicates transpose of a vector such that $\boldsymbol{\pi}$ is a column vector.%
}
	
	We briefly review the post- and pre-order partial likelihood vectors as defined in \cite{ji2020gradients} and use them to derive the partial derivatives.
	For any node $i$ on $\phylogeny$, we divide the observed characters
	$\mathbf{Y} = \{Y_1, Y_2, \dots, Y_{\nTips}\}$
	into two disjoint sets and use them to define the corresponding post- and pre-order partial likelihood vectors for the node.
	We denote the observed characters at the tip nodes that are descendant of node $i$ by $\below{i}$.
	We denote the complement set of observed characters by $\abbove{i} = \mathbf{Y} \setminus \below{i}$.
	We denote the unobserved characters at internal nodes by $\mathbf{y} = \{Y_{N+1}, Y_{N+2}, \dots, Y_{2N-1}\}$. 
	We further denote the branch-specific parameter $\theta_i$ that determines branch-specific $\mathbf{Q}_i$ for branch $i$ and its partial derivative matrix $\dQ{i}$
	such that each entry of $\dQ{i}$ is a partial derivative of the corresponding entry of $\mathbf{Q}_i$ w.r.t.~$\theta_i$.
	Let $\Allparameter = \{\phylogeny, \br{i}, \nodeT{i}, \theta_i, \mathbf{Q}_i; \forall\ i\}$ denote the complete set of all model parameters.
	Finally, we define the length $m$ post-order partial likelihood vector $\p{i}$ of node $i$ at a site as the conditional probability of observing all data at or below node $i$ (i.e.,~$\below{i}$) given the state of node $i$ (i.e.,~$\mathbf{Y}_i$) such that the $s$th element of the post-order partial likelihood vector is $(\p{i})_s = \cDensity{\below{i}}{{Y}_i = s}$.
	Similarly, we define the pre-order partial likelihood vector $\q{i}$ of node $i$ as the joint probability of observing the rest of the data (i.e.,~$\abbove{i}$) and the state of node $i$ (i.e.,~$\mathbf{Y}_i$) such that the $s$th element of the pre-order partial likelihood vector is $(\q{i})_s = \jDensity{\abbove{i}}{\mathbf{Y}_i = s}$.
	We set the pre-order partial likelihood vector at the root node to be the state distribution at the root node (i.e.,~$\q{2\nTips-1} = \boldsymbol{\pi}$).

	\subsection{Likelihood}
	
	If the states of the internal nodes are all observed, one could write out the joint likelihood by
\begin{equation}
	\label{Eq:joint_likelihood}
	\begin{aligned}
		{\Pr{(\mathbf{Y}, \mathbf{y})}}
		&=
		\Pr{(Y_{2N-1})} \prod\limits_{j=1}^{2N-2} {\cDensity{Y_j}{Y_{\parent{j}}}}.\\
	\end{aligned}
\end{equation}
	However, the data likelihood is the probability of the observed discrete characters at the tip nodes by marginalizing over all possible latent states at the internal nodes
	\begin{equation}
		\label{Eq:likelihood}
		\begin{aligned}
			\Pr{(\mathbf{Y})} &=
			\sum\limits_{{Y_{N+1}}}\sum\limits_{{Y_{N+2}}} \ldots \sum\limits_{{Y_{2N-1}}} {\Pr{(\mathbf{Y}, \mathbf{y})}}.\\
		\end{aligned}
	\end{equation}

        We omit the conditioning on the parameters $\Allparameter$ above and in later derivations for ease of notation. 
	\cite{ji2020gradients} derived the calculations of the post- and pre-order partial likelihood vectors.
	We summarize the updates of the partial likelihood vectors here with more detailed calculations of the transition probability matrix to later apply them for the calculation of the derivatives w.r.t.~substitution parameters.
	\myedit{pFiveJ}{%
	We initialize the post-order partial likelihood vectors at the tip nodes by the corresponding observed sequence state at a site such that  $\cDensity{\below{i}}{Y_i=s} = \cDensity{Y_i}{Y_i=s} = \mathds{1}_{\{Y_i=s\}}$ for $i = 1, 2, \ldots, \nTips$.%
}
	One can modify the post-order partial likelihood vector to account for partially observed and missing data at the tip node \citep{Felsenstein1981}.
	The updates of post-order partial likelihood vectors occur through a post-order traversal where descendant nodes on the phylogenetic tree are visited before their parent nodes.
	More specifically, the update of the post-order partial likelihood vector $\p{k}$ at internal node $k$ with two descendant nodes $i$ and $j$ (i.e.,~$\parent{i} = \parent{j} = k$) where $\p{i}$ and $\p{j}$ are available falls out as
	\begin{equation}
		\label{Eq:postOrderPartialUpdate}
		\p{k}=\Ptr{i} \p{i} \odot \Ptr{j} \p{j},
	\end{equation}
	\myedit{elementWiseProduct}{%
	where $\circ$ denotes the element-wise product (i.e.,~Hadamard product).%
}
	After finishing the post-order traversal, the data likelihood is
	\begin{equation}
		\label{Eq:likelihood_at_root}
			\Pr{(\mathbf{Y})} = \p{2\nTips - 1}\transpose \boldsymbol{\pi},
	\end{equation}%
	\myedit{secondTranspose}{%
	where $\transpose$ indicates vector (or matrix) transpose.%
}
	Similarly, we initialize the pre-order partial likelihood vector at the root node by the state distribution at the root (i.e.,~$\q{2\nTips - 1} = \boldsymbol{\pi}$)
	and update the pre-order partial likelihood vectors through a pre-order traversal that visits nodes on the phylogenetic tree in a parent-node-first manner in the reverse order of a post-order traversal.
	This way, the update of the pre-order partial likelihood vector $\q{i}$ at internal node $i$ with its sister node $j$ and parent node $k$ becomes
	\begin{equation}
		\q{i} = \Ptr{i}\transpose \left[ \q{k} \circ \left( \Ptr{j} \p{j} \right) \right],
	\end{equation}
	where transition probability matrices $\Ptr{i}$ and $\Ptr{j}$ are available in the post-order traversal when calculating the likelihood.
	\myedit{anySingleNode}{%
	With the pre- and post-order partial likelihood vectors, one can rewrite Equation~\ref{Eq:likelihood} by marginalizing over the states of any single node $i$ such that%
}
	\begin{equation}
		\label{Eq:likelihood_over_one}
		\begin{aligned}
			\Pr{(\mathbf{Y})} &=
			\sum\limits_{{Y_{i}}} {\Pr{(\mathbf{Y}, Y_i)}}\\
			&= 	\sum\limits_{{Y_{i}}} {\cDensity{\below{i}}{Y_i}  \Pr{(\abbove{i}, Y_i)}}\\
			&=  \p{i}\transpose \q{i}.\\
	\end{aligned}
	\end{equation}
	Equation~\ref{Eq:likelihood_over_one} forms the basis for the gradient algorithm and holds for all nodes (i.e.,~$i = 1, 2, \dots, 2\nTips - 1$).
	The likelihood calculation at the root after finishing the post-order traversal (i.e.,~Equation~\ref{Eq:likelihood_at_root}) is a special case for $i = 2\nTips - 1$.

\begin{figure} 
		\begin{center}
			  \includegraphics[width=0.4\textwidth]{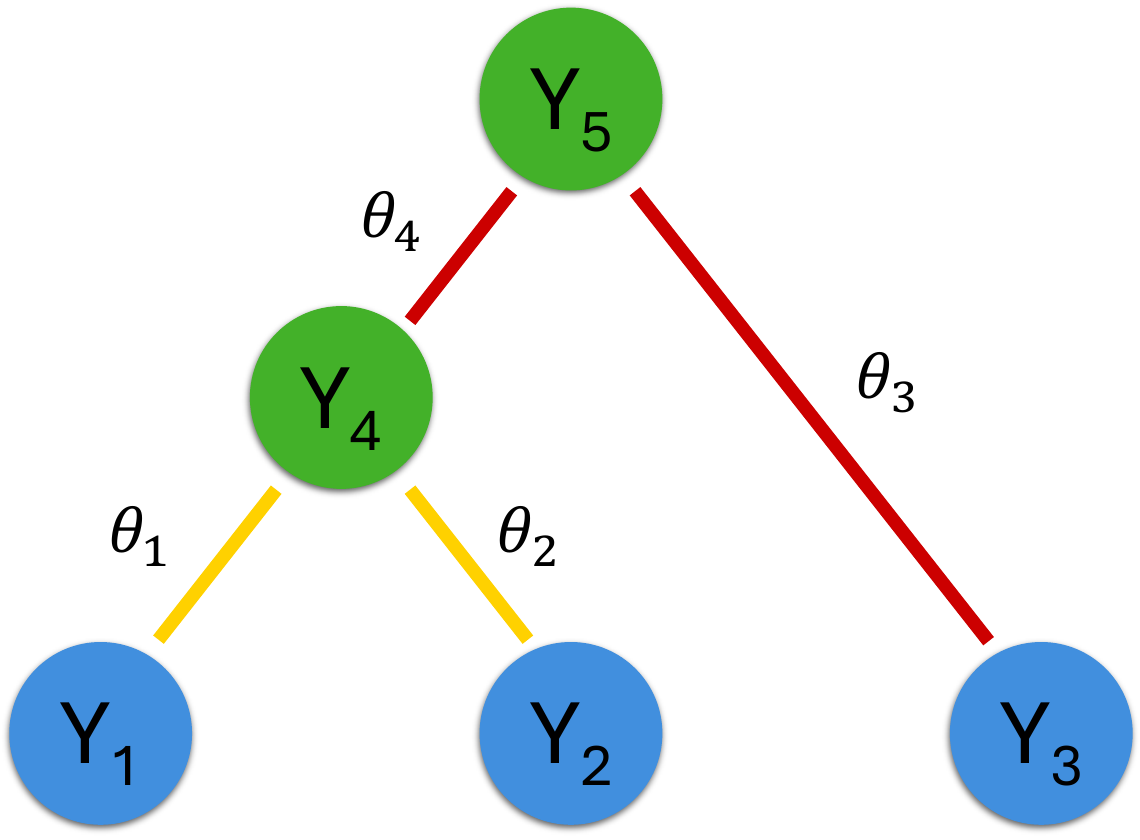}
			\caption{Example of a $3$-taxon tree.
				Sequence states $\mathbf{Y} = ({Y_1}, {Y_2}, {Y_3})\transpose$ are observed data at the tips of the tree.
				The latent states ${Y_4}$ and ${Y_5}$ are unobserved at internal nodes of the tree.
				For node $4$, its post-order partial likelihood vector contains the conditional probability of 
				$ \cDensity{\below{4}}{Y_4}  = \cDensity{Y_1, Y_2}{Y_4}$ and
				its pre-order partial likelihood vector contains the joint probability of
				$ \jDensity{\abbove{4}}{Y_4}  = \jDensity{Y_3}{Y_4}$.
				We further color the branches to show the update of the two partial likelihood vectors at internal node $4$ such that
				yellow branches correspond to the update of the post-order partial likelihood vectors
				and
				red branches correspond to the update of the pre-order partial likelihood vectors.}
			\label{fig:exampleTree}
		\end{center}
	\end{figure}

	Figure~\ref{fig:exampleTree} illustrates these quantities using a 3-taxon tree example.
	The observed data in Figure~\ref{fig:exampleTree} are $\mathbf{Y} = \{{Y_1}, {Y_2}, {Y_3}\}$.
	One obtains the likelihood of the observed data by marginalizing over $\mathbf{y}=\{{Y_4}, {Y_5}\}$.
	Two possible post-order traversals for the example tree in Figure~\ref{fig:exampleTree} are
	 $1 \to 2 \to 4 \to 3 \to 5$ and $1 \to 2 \to 3 \to 4 \to 5$.
	 The reverse of these post-order traversals form the pre-order traversals.
	 In the post-order traversal, one calculates the transition probability matrices at branches $1$ to $4$ by $\Ptr{i} = \mathbf{U}_i e^{\boldsymbol{\Lambda}_i \bl{i}}\mathbf{U}_i^{-1}$, $i = 1, 2, 3, 4$ where the eigen decomposition is either analytically (e.g.,~\cite{tamura1993estimation} model and its nested special cases) or numerically available.
	 The updates of post-order partial likelihood vectors become $\p{4} = \Ptr{1}\p{1} \circ \Ptr{2} \p{2}$ and $\p{5} = \Ptr{4}\p{4} \circ \Ptr{3} \p{3}$ where post-order partial likelihood vectors ($\p{1}$, $\p{2}$, $\p{3}$) are formed according to the observed sequence state at the corresponding tip nodes.
	 For the pre-order traversal, one initializes the pre-order partial likelihood vector at the root node by setting $\q{5} = \boldsymbol{\pi}$.
	 The updates of pre-order partial likelihood vectors become $\q{4} = \Ptr{4}\transpose \left[ \q{5} \circ \left( \Ptr{3} \p{3} \right) \right]$, $\q{3} = \Ptr{3}\transpose \left[ \q{5} \circ \left( \Ptr{4} \p{4} \right) \right]$, $\q{2} = \Ptr{2}\transpose \left[ \q{4} \circ \left( \Ptr{1} \p{1} \right) \right]$ and $\q{1} = \Ptr{1}\transpose \left[ \q{4} \circ \left( \Ptr{2} \p{2} \right) \right]$.

	\subsection{Gradient}
	
	\cite{ji2020gradients} detail 
	the calculation of the derivative w.r.t.~branch lengths, hence we will here focus on the calculation of the derivative w.r.t.~parameters within the infinitesimal generator matrix.
	With the likelihood expanded at node $i$ as in Equation~\ref{Eq:likelihood_over_one}, we derive the corresponding scalar component of gradient vector 
	which is the first order derivative w.r.t.~parameter $\theta_i$ (i.e.,~$\left[\nabla {\lnP{\mathbf{Y}}}\right]_i = \gradient{\theta_i}{\lnP{\mathbf{Y}}}$).
	To do this, we adopt a ``spectral representations'' approach to calculate the first directional derivative of the matrix exponential as in \cite{najfeld1995derivatives}.
	We first derive the derivative of the transition probability matrix $\Ptr{i}$ w.r.t.~$\theta_i$,
	\begin{equation}
		\label{Eq:derivPtr}
		\begin{aligned}
			\frac{\partial \Ptr{i}}{\partial \theta_i}
			&= \frac{\partial e^{\mathbf{Q}_i \bl{i}}}{\partial \theta_i}\\
			&=\mathbf{U}_i \left[ \bar{\mathbf{Q}}_i \circ \boldsymbol{\Phi}_i (\bl{i})\right] \mathbf{U}_i^{-1},\\
		\end{aligned}
	\end{equation}
where $\bar{\mathbf{Q}}_i  = \mathbf{U}_i^{-1} \dQv{i} \mathbf{U}_i $.
	 The matrix $\boldsymbol{\Phi}_i(\bl{i})$ is symmetric with entries
	\begin{equation}
		\label{eq:phi}
		\{\boldsymbol{\Phi}_i(\bl{i})\}_{jk} = \{\boldsymbol{\Phi}_i(\bl{i})\}_{kj} = \left\{ \begin{array}{ll}
			(e^{\bl{i}\lambda_{ij}} - e^{\bl{i}\lambda_{ik}}) / (\lambda_{ij} - \lambda_{ik}) & \mbox{if } \lambda_{ij} \ne \lambda_{ik}\\
				\bl{i} e^{\bl{i}\lambda_{ij}} & \mbox{if } \lambda_{ij} = \lambda_{ik},
				\end{array} \right.
	\end{equation}
	and the derivative of the pre-order partial likelihood vector $\q{i}$ w.r.t.~$\theta_i$ becomes
	\begin{equation}\begin{aligned}
		\label{Eq:preOrderPartialGradient}
		\frac{\partial \q{i}}{\partial \theta_i}
		&= \gradient{\theta_i}{\left\{ \Ptr{i}\transpose \left[ \q{k} \circ \left( \Ptr{j} \p{j} \right) \right] \right\}}
		=  \left( \gradient{\theta_i}{e^{\mathbf{Q}_i \bl{i}}} \right)\transpose \left[ \q{k} \circ \left( \Ptr{j} \p{j} \right) \right] \\
		&=  \left\{ {e^{\mathbf{Q}_i \bl{i}}e^{-\mathbf{Q}_i \bl{i}} \cdot   \mathbf{U}_i \left[ \bar{\mathbf{Q}}_i \circ \boldsymbol{\Phi}_i(\bl{i})\right] \mathbf{U}_i^{-1} } \right\}\transpose \left[ \q{k} \circ \left( \Ptr{j} \p{j} \right) \right] \\                
		&=  \left\{ {e^{\mathbf{Q}_i \bl{i}}  \left[\mathbf{U}_i e^{-\boldsymbol{\Lambda}_i \bl{i}}\mathbf{U}_i^{-1} \right] \cdot \mathbf{U}_i \left[ \bar{\mathbf{Q}}_i \circ \boldsymbol{\Phi}_i(\bl{i})\right] \mathbf{U}_i^{-1} } \right\}\transpose \left[ \q{k} \circ \left( \Ptr{j} \p{j} \right) \right] \\
		&= \left\{ \mathbf{U}_i \left[ \bar{\mathbf{Q}}_i \circ \left( e^{-\boldsymbol{\Lambda}_i \bl{i}} \mathbf{\Phi}_i(\bl{i}) \right)  \right] \mathbf{U}_i^{-1} \right\}  \transpose{\left\{ \Ptr{i}\transpose \left[ \q{k} \circ \left( \Ptr{j} \p{j} \right) \right] \right\}},\\
		&= \left\{ \mathbf{U}_i \left[ \bar{\mathbf{Q}}_i \circ \mathbf{\Psi}_i(\bl{i}) \right] \mathbf{U}_i^{-1} \right\}  \transpose \q{i},\\
	\end{aligned}\end{equation}
	where matrix $\mathbf{\Psi}_i(\bl{i}) = e^{-\boldsymbol{\Lambda}\bl{i}} \boldsymbol{\Phi}_i(\bl{i})$ is symmetric with entries
	\begin{equation}
		\label{eq:Psi}
			\{\mathbf{\Psi}_i(\bl{i})\}_{jk} = \{\mathbf{\Psi}_i(\bl{i})\}_{kj}  = \left\{ \begin{array}{ll}
			\left[1 - e^{\bl{i}(\lambda_{ik} - \lambda_{ij})}\right] / (\lambda_{ij} - \lambda_{ik}) & \mbox{if } \lambda_{ij} \ne \lambda_{ik}\\
			\bl{i} & \mbox{if } \lambda_{ij} = \lambda_{ik}.\\
		\end{array} \right.
	\end{equation}
	Finally, the derivative of the log likelihood w.r.t.~$\theta_i$ falls out as
	\begin{equation}
	\label{Eq:singleRateCaseGradient}
	\begin{aligned}
		\gradient{\theta_i}{\lnP{\mathbf{Y}}}
		&=\gradient{\theta_i}{\left[\p{i}\transpose \q{i} \right]} \bigg/ \Pr(\mathbf{Y})\\
		&=\p{i}\transpose \frac{\partial \q{i}}{\partial \bl{i}} \bigg/ \Pr(\mathbf{Y})\\
		&= \q{i}\transpose  \left\{ \mathbf{U}_i \left[ \bar{\mathbf{Q}}_i \circ \mathbf{\Psi}_i(\bl{i}) \right] \mathbf{U}_i^{-1} \right\} \p{i} \bigg/ \Pr(\mathbf{Y}).\\
	\end{aligned}
	\end{equation}

	\subsection{Likelihood and gradient with substitution rate heterogeneity}
	\label{sec:rate_heterogenity}
	
	Equation~\ref{Eq:singleRateCaseGradient} does not consider substitution rate heterogeneity across sites.
	\myedit{HMM}{%
	A popular approach to model the substitution rate heterogeneity across sites is by using a mixture model 
	where the substitution rate of a site belongs to one of multiple rate categories \citep{yang1994maximum}.%
}
	For discrete rate category $l$ with rate $\sr{l}$, the transition probability matrix for branch $k$ of rate category $l$ becomes $\Ptr{k|\sr{l}} = e^{\mathbf{Q}_k \bl{k} \sr{l}}$.
	\myedit{HMMTwo}{%
	As in mixture models, the observed data likelihood marginalizes over all rate categories and becomes the weighted sum of the conditional likelihood (of each rate category):%
}
	%
	\begin{equation}
		\label{Eq:discreteRateLikelihood}
		\begin{aligned}
			\Pr(\mathbf{Y})
			&= \sum_{\sr{l}} \cDensity{\mathbf{Y}}{\sr{l}} \Pr(\sr{l})\\
			&= \sum_{\sr{l}} \p{k|\sr{l}}\transpose \q{k|\sr{l}}  \Pr(\sr{l}),\\
		\end{aligned}
	\end{equation}
	where $\p{k|\sr{l}}$ and $\q{k|\sr{l}}$ are the corresponding post- and pre-order partial likelihood vectors at
	node $k$ for rate category $l$.
	\cite{ji2020gradients} detailed their updates.
	Similarly, the numerator and denominator of Equation~\ref{Eq:singleRateCaseGradient} become weighted sums in the rate heterogeneous case:
	%
	\begin{equation}\label{Eq:GradientWithGamma}
		\begin{aligned}
			\gradient{\bl{i}}{\lnP{\mathbf{Y}}}
			&=
			\sum\limits_{\sr{l}} \sr{l}  \q{i|\sr{l}}\transpose  \left\{ \mathbf{U}_i \left[ \bar{\mathbf{Q}}_i \circ \mathbf{\Psi}_i(\bl{i}\sr{l}) \right] \mathbf{U}_i^{-1} \right\} \p{i|\sr{l}} \Pr(\sr{l}) \bigg/ {\Pr(\mathbf{Y})}.\\
		\end{aligned}
	\end{equation}
	%
	Similar to the calculation of the gradient w.r.t.~branch lengths \citep{ji2020gradients}, Equation~\ref{Eq:preOrderPartialGradient} and Equation~\ref{Eq:GradientWithGamma} show that we can reuse the post- and pre-order partial likelihood vectors $\p{i}$, $\q{i}$ by constructing a modified partial derivative matrix ($\mathbf{U}_i \left[ \bar{\mathbf{Q}}_i \circ \mathbf{\Psi}_i(\bl{i}) \right] \mathbf{U}_i^{-1}$ for rate homogeneity and $\mathbf{U}_i \left[ \bar{\mathbf{Q}}_i \circ \mathbf{\Psi}_i(\bl{i} \sr{l}) \right] \mathbf{U}_i^{-1}$ for  rate heterogeneity across sites) at node $i$ for calculating the partial derivative of branch $i$.
	Since the modified partial derivative matrix construction is only branch-dependent, we can calculate these branch-specific derivatives together with the update of the pre-order partial likelihood vectors.
	The derivative w.r.t.~different substitution parameters is associated with different partial differential matrices and their calculation can happen at the same time using the same set of post- and pre-order partial likelihood vectors.
	This procedure gives us the gradient vector of all partial derivatives w.r.t.~branch $1$, $2$, \ldots, $2\nTips - 2$ in one single pre-order traversal in linear-time.
	
	\subsection{Example analytic derivatives w.r.t.~substitution parameters}
	\label{sec:substitution_model}
	
	We explore two specific branch-specific substitution models to learn the evolutionary dynamics of selection pressure and mutational changes.
	We refer to the $4$-state nucleotide substitution  model as the HKY+APOBEC model that extends the HKY model \citep{hasegawa1985dating} to account for excess $C$ to $T$ and $G$ to $A$ transitions due to effect of host APOBEC proteins in mpox evolution \citep{o2023apobec3}.
        The other is a $61$-state codon substitution model denoted as $F1 \times 4MG + \kappa + \omega$ by the codeml software \citep{yang2007paml} with branch-specific $\omega$ parameters.
	We derive the analytic derivative forms w.r.t.~APOBEC-effect parameter $\tau$ and the $dN/dS$ parameter $\omega$ of these two substitution models respectively using Equation~\ref{Eq:singleRateCaseGradient}.
	For simplicity, we drop the subscript $i$ and derive the derivative form without rate heterogeneity that is the equivalent form for one single rate category (when considering rate heterogeneity).
	
	The HKY + APOBEC model has the off-diagonal instantaneous rate $Q_{ij}$ from nucleotide $i$ to $j$ be
	\begin{equation}
		\label{eq:HKY}
		Q_{ij} = \left\{
			\begin{tabular}{ll}
				$\pi_j$ & {for a transversion}\\
				$\pi_j \kappa$ & {for a non-APOBEC transition}\\
				$\pi_j \kappa \tau$ & for an APOBEC transition,\\
			\end{tabular}
		\right.
	\end{equation}
	where APOBEC transitions correspond to $(i, j) = (C, T)$ and $(i, j) = (G, A)$ (see Section~\ref{sec:mpox} for more details) and the APOBEC-parameter $\tau > 0$ captures the multiplicative effect such that $\tau > 1$ indicates excess APOBEC transition rates.
	The rows of $\mathbf{Q}$ sum to $0$ so that the diagonal entries of $\mathbf{Q}$ are the negative of the row sum of off-diagonal entries ($Q_{i i} = - \sum\limits_{j \neq i} Q_{i j}$).
	The off-diagonal entry of differential matrix $\dQ{}$ is then
		\begin{equation}
		\label{eq:HKY_derivative}
		\left(\dQv{}\right)_{i j} = \left\{
		\begin{tabular}{ll}
			$\pi_j \kappa$ & for an APOBEC transition\\
			$0$ & otherwise,\\
		\end{tabular}
		\right.
	\end{equation}
	where row sums of $\dQ{}$ are $0$.

	The  $F1 \times 4MG + \kappa + \omega$ codon substitution model has the instantaneous rate $Q_{i j}$ from codon triplet $i$ to $j$ be $0$ if $i$ and $j$ differ in more than one of their three positions. 
	If $i$ and $j$ differ in exactly one nucleotide that has type $h$ ($h \in \{A, C, G, T\}$) in codon $j$, the instantaneous rate becomes
		\begin{equation}
		\label{eq:codon}
		Q_{i j} = \left\{
		\begin{tabular}{ll}
			$\pi_h$ & {for a synonymous transversion}\\
			$\pi_h \kappa$ & {for a synonymous transition}\\
			$\pi_h \omega$ & {for a nonsynonymous transversion}\\
			$\pi_h \kappa \omega$ & {for a nonsynonymous transition,}\\
		\end{tabular}
		\right.
	\end{equation}
	and the differential matrix $\dQ{}$ has off-diagonal entries
		\begin{equation}
	\label{eq:codon_derivative}
	\left(\dQv{}\right)_{i j} = \left\{
	\begin{tabular}{ll}
		$\pi_h$ & {for a nonsynonymous transversion}\\
		$\pi_h \kappa$ & {for a nonsynonymous transition}\\
		$0$ & otherwise.\\
	\end{tabular}
	\right.
\end{equation}
Similar to the case of the HKY + APOBEC model, $\mathbf{Q}$ and $\dQ{}$ of the $F1 \times 4MG + \kappa + \omega$ model have row sums of $0$.
While Equations~\ref{eq:HKY_derivative} and \ref{eq:codon_derivative} only form the differential matrix, one can easily construct the $\boldsymbol{\Psi}$ matrix through Equation~\ref{eq:Psi} and $\bar{\mathbf{Q}}$ using the eigen decomposition of the instantaneous transition matrices $\mathbf{Q}$.

	

	\subsection{Autocorrelated shrinkage branch-specific substitution parameters}
	We assume the branch-specific substitution parameters $\boldsymbol{\theta} = \{\theta_1, \dots, \theta_{2\nTips - 2} \}$ are autocorrelated.
        Similar to the treatment of branch-specific evolutionary rates in \cite{fisher2023shrinkage},
        we model the autocorrelation in terms of the incremental difference $\phi_i$ between the log-transformed values of branch $i$'s substitution parameter and its parent lineage's substitution parameter: 
	%
	\begin{equation}
		\phi_i = \log \theta_i - \log \theta_{\parent{i}}, \mbox{for } i \in \{1, \dots, 2\nTips - 2\} \mbox{ and let } \theta_{2\nTips - 1} \equiv 1.
	\end{equation}
	Under this parameterization, the increments $\boldsymbol{\phi} = \{\phi_1, \dots, \phi_{2\nTips - 2}\} \in \mathbb{R}^{2\nTips - 2}$ are a linear transformation of $\log \boldsymbol{\theta}$.
	To shrink the total number of substitution parameter changes along the tree, we employ shrinkage priors on $\phi_i$'s such that $\phi_i \overset{iid}{\sim}P_\phi$ and $E(\phi_i) = 0$.
	Typically, $P_\phi$ may follow a Gaussian \citep{thorne1998estimating}, Laplace or Horseshoe distribution \citep{carvalho2010horseshoe}.
	We choose the flexible, heavy-tailed, Bayesian bridge prior \citep{polson2014bayesian} on the increments,
	\begin{equation}
		P_\phi \propto \exp\left\{-\left| \frac{\phi_i}{\mu}\right|^\alpha \right\},
	\end{equation}
	where $\mu > 0$ is termed the ``global scale'' and $\alpha \in (0, 1]$ changes the shape of $P_\phi$ so that smaller $\alpha$ places more mass near zero.
	When $\alpha = 1$,  $P_\phi$ is the Laplace prior.
	When $\alpha$ is closer to $0$, the Bayesian bridge prior approaches the best subset selection when used in a regression setting.
	In both examples, we set $\alpha = 0.9$ to enforce slightly weaker shrinkage compared to \cite{fisher2023shrinkage}.
	Therefore, the joint prior for all increments is the product 		$p(\boldsymbol{\phi}) \propto \prod\limits_{i = 1}^{2\nTips - 2} \exp\left\{ - \left| \frac{\phi_i}{\mu} \right|^\alpha \right\}$.

\newcommand{\position}{\boldsymbol{\theta}}

	\subsection{Hamiltonian Monte Carlo method}
	\label{sec:HMC}
	We briefly review the HMC sampling method.
	HMC has demonstrated effective performance in learning high-dimensional posterior distributions associated with modern phylodynamic models \citep{ji2020gradients,  fisher2021relaxed, fisher2023shrinkage, ji2023scalable, hassler2023data}.
	The linear-time analytical gradient algorithm developed in this manuscript further expands application of HMC.
	HMC is a state-of-the-art Markov chain Monte Carlo (MCMC) method that proposes new values for all parameters with relatively high acceptance rate by exploiting Hamiltonian dynamics \citep{neal2011mcmc}.
	To sample from the posterior distribution $\pi(\boldsymbol{\theta})$, HMC introduces an auxiliary momentum parameter $\momentum$ usually drawn from a multivariate normal distribution with mean $\mathbf{0}$ and variance-covariance matrix $\mass$ that is also referred to as the `mass matrix'  (i.e.,~$\momentum \sim \mathcal{N}(\mathbf{0},\mass)$).
	The auxiliary momentum parameter has `kinetic energy' of $K(\momentum) = \momentum\transpose \mass^{-1} \momentum/2$ (due to the multivariate normal kernel).
	HMC treats the negative log density of $\pi(\boldsymbol{\theta})$ as the `potential energy' $U(\position) = -\log(\pi(\position))$ and the combination of the potential and kinetic energies form the Hamiltonian function $H(\position, \momentum) = U(\position) + K(\momentum)$.
	HMC generates a Metropolis proposal \citep{metropolis53} by simulating Hamiltonian dynamics from the current state $(\position_0, \momentum_0)$ for a pre-defined integration time according to the differential equation:
	\begin{equation}
		\label{eq: HMC}
		\begin{aligned}
			\frac{\textrm{d} \position}{\textrm{d} t} &=  \nabla K(\momentum) = \mass^{-1} \momentum\\
			\frac{\textrm{d} \momentum}{\textrm{d} t} &=  - \nabla U(\position) = \nabla \log \pi(\position),\\
		\end{aligned}
	\end{equation}
	that is often numerically integrated through the {\it leapfrog} method \citep{neal2011mcmc} as an approximation.
	The leapfrog method performs $n$ step discrete integration with a step size of $\epsilon$ such that the total integration time becomes $n\epsilon$.
	For each leapfrog step, one iteratively updates the momentum and position by
	\begin{equation}
		\label{eq:leapfrog}
		\begin{aligned}
			\momentum_{t + \epsilon/2} &= \momentum_{t} + \frac{\epsilon}{2} \nabla \log \pi(\position_t)\\
			\position_{t + \epsilon} &= \position_{t} + \epsilon \mass^{-1} \momentum_{t + \epsilon/2}\\
			\momentum_{t + \epsilon} &= \momentum_{t + \epsilon/2} + \frac{\epsilon}{2} \nabla \log \pi(\position_{t + \epsilon}) \, .
		\end{aligned}
	\end{equation}
	The leapfrog integration starts from time $t = 0$ and ends at $t = n\epsilon$ where the ending position $\position_{n\epsilon}$ is the proposed state.
	The proposed state is then accepted with probability
	\begin{equation}
		\min \left\{1, \exp \left[{H(\position_0, \momentum_0) - H(\position_{n\epsilon}, \momentum_{n\epsilon})} \right] \right\},
	\end{equation}
	that depends on how well the numerical integration preserves the Hamiltonian energy.

	The geometric structure of the posterior distribution substantially influences the computational efficiency of HMC.
	Specifically, when individual parameters exhibit varying scales within the posterior, neglecting this structural variation can lead to a significant reduction in HMC efficiency \citep{neal2011mcmc, ji2020gradients, fisher2023shrinkage, ji2023scalable}.
	To account for structural variability in the posterior, we adopt the preconditioning mass matrix informed by the absolute value of the diagonal Hessian of the log-prior density as developed in \cite{fisher2023shrinkage}.

	\section{Materials and Methods}

	\subsection{Sequence data sets}

	We examine the mutational dynamics in the molecular evolution of the 2022 mpox epidemic 
	 in North America \citep{o2023apobec3, paredes2024underdetected} and the classic tumor suppressor BRCA1 genes from primates \citep{yang2002codon}.

	\paragraph{MPXV}
	\label{sec:mpox}
	Mpox is a viral zoonotic disease caused by the mpox virus (MPXV) endemic to West and Central Africa.
	MPXV is a double-stranded DNA virus and was first discovered in 1958 in Copenhagen \citep{cho1973monkeypox}.
	In July 2022, the World Health Organization declared mpox a public health emergency of international concern.
	The DNA sequences for the first cases from the 2022 mpox epidemic shared $42$ nucleotide differences from the closest MPXV sequences sampled prior to the outbreak. 
	Almost all of these mutations are typical of the activity of APOBEC3 deaminases, which are host enzymes that play a role in antiviral defense.
	Assuming APOBEC3 editing occurs at a higher rate during human MPXV infection than in the previous host, 
	\cite{o2023apobec3} developed a dual-process molecular clock model to account for possible elevated APOBEC3 mutation rates and estimated that MPXV had been circulating in humans since 2016.
	Interestingly, most of the observed nucleotide changes associated with the MPXV sequences from the 2022 epidemic appear to be a particular dinucleotide change from TC $\to$ TT and its reverse complement, GA $\to$ AA.
	In light of this specific dinucleotide mutation pattern, we extend the HKY model with a new parameter to account for possible excess C $\to$ T and G $\to$ A mutations as illustrated in Equation~\ref{eq:HKY}.
	We estimate the dynamic of APOBEC3 mutation rates in recent MPXV evolution. 
	The MPXV data example consists of $138$ genomes with an alignment of $197,209$ nucleotides.

	\paragraph{BRCA1 gene}
	The BRCA1 tumor suppressor gene contributes to preserving genomic integrity by participating in recombinational and transcription-coupled DNA repair, as well as regulating transcription.
	Mutations in the BRCA1 gene are associated with a higher risk of developing breast cancer in women.
		In their landmark study, 
		\cite{yang2002codon} studied the protein coding sequences of BRCA1 in primates using a branch-specific $\omega$ model and observed an elevated estimate of $dN/dS$ values (i.e.,~the $\omega$ parameter) in the clade of human and chimpanzee that suggested positive selection.
		We re-examine this classic data set that contains $7$ 
		ingroup primate species (human, chimpanzee, gorilla, orangutan, macaca, howler monkey and bushbaby) and an outgroup species (flying lemur) with an alignment of $1,123$ codons.

	\subsection{Other models and priors}
	We employ branch-specific substitution processes as detailed in Section~\ref{sec:substitution_model} in our analyses.
	We place Bayesian bridge shrinkage priors on the log difference of the branch-specific substitution parameter of a branch and that of its parent branch with  the global scale $\mu = 1$ and exponent $\alpha = 0.9$ for both analyses.
	We pick this specific exponent value to shrink the total number of change-points  in the posterior across the tree to around $2$  for the MPXV data example 
	and use the same exponent value for the BRCA1 example. 
	We place a normal prior with mean $0$ and variance $1$ on the log of the kappa parameter in the HKY model for the MPXV analysis and an exponential prior with mean $1$ on the kappa parameter for the BRCA1 analysis. 
	\myedit{kappa}{%
	The different prior choices for the kappa parameter stem from our modeling and implementation choices.  On one hand, we model instantaneous transition rates of the HKY + APOBEC process in the log space to more conveniently compose the differing effects while retaining positivity; whereas in the $F1 \times 4MG + \kappa + \omega$ codon substitution model, we model the instantaneous transition rates directly in their original space.%
}
	We place a coalescent prior with exponentially growing effective population size on ingroup taxa and constant effective population size on outgroup taxa on the tree topology for the MPXV analysis (please see the BEAST XML files for details of ingroup and outgroup taxa) and a fixed species tree topology for the BRCA1 analysis. 
	We employ a random-effects molecular clock model as described in \cite{ji2020gradients} for the MPXV analysis and a fixed evolutionary rate of $1$ for the BRCA1 analysis.
	We employ a discretized gamma distribution with $4$ rate categories for modeling among-site rate heterogeneity \citep{yang1994maximum} for both analyses and place an exponential prior with mean $0.5$ on its shape parameter.

	\subsection{Implementation}

	The gradient algorithm derived in Equation~\ref{Eq:preOrderPartialGradient} and Equation~\ref{Eq:GradientWithGamma} uses the same BEAGLE \citep{ayres2019beagle} infrastructure with a CPU implementation as described in \cite{ji2020gradients} and a multi-core GPU implementation as in \cite{gangavarapu2024many}.
	We implement the construction of the modified partial differential matrix within the development branch ``hmc-clock'' of  BEAST X \citep{suchard2018, baele2025} for the demonstrations in this paper.
	We provide instructions, commit tags of BEAGLE and BEAST X software packages, and the BEAST XML files for reproducing these analyses on GitHub at \href{https://github.com/suchard-group/BranchSpecificSupplementary}{https://github.com/suchard-group/BranchSpecificSupplementary}.

	\section{Results}
	We demonstrate the computational efficiency gains of our linear-time gradient algorithm for inferring the branch-specific substitution parameters using two empirical data examples of different sizes.
        The first dataset (MPXV) includes a large sample of $138$ complete genomes, 
        while the second (BRCA1) includes a small sample of $8$ sequences. 
	For each dataset, we present results from both optimization (via L-BGFS) and Bayesian posterior sampling (via MCMC).
        We compare performance of the analytic gradients versus central-difference numerical gradients by applying them separately in the gradient-based L-BFGS optimization routine (see e.g.,~\cite{dennis1996numerical}). 
        For posterior sampling, we assess the efficiency of HMC relative to single-variable Metropolis-Hastings proposals.

	\subsection{Optimization}
	We obtain the maximum likelihood estimate (MLE) of the branch-specific substitution parameters conditional on all other parameters via the L-BFGS optimization algorithm for both datasets.
	We compare the performance of our analytic gradient method with an often-used central finite difference numerical  scheme.
	The numerical  scheme calculates the derivative of one branch-specific substitution parameter via two likelihood evaluations and therefore has a complexity of $O(\nTips^2)$ for the gradient w.r.t.~all $O(\nTips)$ dimensions.
	On the other hand, our analytic approach scales linearly w.r.t.~number of sequences and only requires $O(\nTips)$ computation to achieve the same gradient.
	Table~\ref{tab:MLE_Speedup} provides a summary of the comparison, highlighting the significant performance improvement of our analytic method across the two datasets.
	In terms of overall computational time for the MLE estimation procedure, the analytic gradient method achieves a performance advantage, offering approximately a 84.9-fold speedup for the MPXV example and a 2.4-fold speedup for the BRCA1 example compared to the central finite difference numerical gradient method.
	When averaged over each iteration of the MLE estimation procedure, the analytic gradient method achieves an 126.1-fold speedup for the MPXV example and a 7.0-fold speedup for the BRCA1 example compared to the central finite difference numerical gradient method. 
	The analytic gradient also achieves higher ending maximum log likelihood values in both examples.

        %
	\begin{table}
		\caption{
			Maximum likelihood estimate (MLE) inference efficiency using two optimization methods: our proposed gradient method (Analytic) and a central finite difference numerical scheme (Numerical).
			For each example and method, we report the maximized log likelihood, total time to complete MLE inference, as well as the number of iterations required for optimization.
			Our proposed method yields a higher maximized log likelihood value with at least 7-fold increase in performance per iteration across the entire inference. 
		}
		\label{tab:MLE_Speedup}
		\centering
		\resizebox{\textwidth}{!}{\begin{tabular}{lcrrrrrrrr}
				\toprule
				& &  \multicolumn{3}{c}{Analytic} & \multicolumn{3}{c}{Numerical } & \multicolumn{2}{c}{Speedup}\\
				\cmidrule(lr){3-5} \cmidrule(lr){6-8} \cmidrule(lr){9-10}
				\multicolumn{1}{c}{Example} & \multicolumn{1}{c}{$N$} & \multicolumn{1}{c}{lnL} & Time(s) & Iterations   & \multicolumn{1}{c}{ lnL} & Time(s) & Iterations & \multicolumn{1}{c}{per Iteration} & \multicolumn{1}{c}{Total} \\
				\hline
				MPXV      & 138 & -272882.9253 & 4.3 & 237 & -272889.7849 & 1349.9 & 590 & 126.1$\times$ & 84.9$\times$ \Bstrut\\
				BRCA1       & 8 & -9358.3565 & 15.7  & 223 &  -9358.3565 & 37.8  & 77 & 7.0$\times$ & 2.4$\times$ \Tstrut\\
				\bottomrule
		\end{tabular}}
	\end{table}
	\subsection{Posterior inference}

	\paragraph{MPXV}
	We infer the posterior distribution of all parameters of the model.
        For the MPXV analysis, this includes the substitution parameters, internal node heights, tree topology, shape of the discrete gamma rate heterogeneity model, and evolutionary rates.
	We run two independent MCMC analyses with different random seed numbers to ensure convergence in BEAST X \citep{suchard2018, baele2025} using BEAGLE \citep{ayres2019beagle, gangavarapu2024many} for parallel computation using GPU (for the full analysis) and CPU with Streaming SIMD Extensions (SSE; for benchmarking).
	Our analysis estimates the tree-wise (fixed-effect) mean rate with posterior mean $1.24$ ($95\%$ Bayesian credible interval: $0.93$, $1.59$) $\times 10^{-5}$ substitutions per site per year and an estimated variability characterized by the scale parameter of the lognormal distributed branch-specific random-effects with posterior mean $1.80$ ($1.26$, $2.41$).
	Our estimated evolutionary rate (over the entire sequence) is in between the estimated values of APOBEC3 and non-APOBEC3 evolutionary rates by \cite{o2023apobec3}.
	Figure~\ref{fig:MPXV} shows the maximum clade credibility tree of the MPXV example.
	Our analysis revealed a clear acceleration of the APOBEC effect in the ingroup samples after the split indicated by the blue star branch.
	The estimated time of the split has $95\%$ posterior estimate of 2012.3 (2010.8, 2016.0). 	
%

\begin{figure}
	\begin{center}
		\includegraphics[width=1\textwidth]{APOBEC\_tree\_v2.pdf}
	\end{center} \vspace{-0.8cm}
	\caption{
		Maximum clade credibility tree of the MPXV analysis.
		The dataset consists of $138$ sequences of the mpox virus. 
		Branches are color-coded by the posterior means of the branch-specific APOBEC rate $\tau$.
		The blue star indicates the branch after which the acceleration of APOBEC effect started.
	}
	\label{fig:MPXV}
\end{figure}
	\paragraph{BRCA1}
	We fix the tree topology and infer a rooted tree without imposing any molecular clock models for the BRCA1 analysis such that branch lengths are in the unit of expected codon substitutions per site (i.e.,~equivalent as inferring a time-calibrated tree with fixed molecular clock rate of $1$). 
	Our analysis estimates the transition to transversion ratio parameter $\kappa$  with posterior mean $4.6$ ($4.0$, $5.3$).
	Figure~\ref{fig:BRCA1} shows the estimated posterior means of the branch-specific $dN/dS$ parameter $\omega$ on each branch.
	Our analysis reaffirms the elevated values of the $dN/dS$ ratios leading to the clade of human and chimpanzee as originally revealed by \cite{yang2002codon} while allowing each branch to have its own parameter without any prior allocations on change-point locations.
%
%
%
\begin{figure}
	\begin{center}
		\includegraphics[width=1\textwidth]{BRCA1\_v2.pdf}
	\end{center} \vspace{-0.8cm}
	\caption{
		Maximum clade credibilty tree of the BRCA1 analysis.
		Numbers on branches represent the posterior mean estimates for the branch-specific $\omega$ parameter with numbers in the parenthesis showing the $95\%$ Bayesian credible intervals.
		Branches are color-coded according to the posterior means of the branch-specific $\omega$ parameter.
		Taxon silhouettes were sourced from PhyloPic (\href{https://www.phylopic.org/permalinks/ca29b0ac3b1f3a64f52dd1a92730f45f7c728074b3a5a03c42a027d5671cca78}{phylopic.org}); see link for attribution.
	}
	\label{fig:BRCA1}
\end{figure}

	\subsection{Sampling efficiency}
	We infer the conditional posterior distribution of all branch-specific substitution parameters using an HMC and a univariate Metropolis-Hastings transition kernel for the MPXV and BRCA1 examples.
	We refer to the univariate transition kernel as `univariate'.
	The `univariate' updates propose new values for a single dimension in the high-dimensional branch-specific substitution parameter at a time where the HMC updates propose new values for all dimensions simultaneously by integrating the Hamiltonian dynamics (see Section~\ref{sec:HMC}).
	We compare the efficiency of these two transition kernels by their ability to accumulate effective sample size (ESS) per unit time for estimating the high-dimensional branch-specific substitution parameters.
	\myedit{ESS}{%
	We calculate the ESS values using the ``coda'' package \citep{coda} in R \citep{R} and provide our R script together with all log files in the Supplementary Material.%
}
	For each analysis, we adjust the MCMC iterations such that the chains with different transition kernels run for comparable wall time.
	Specifically, we run the MPXV experiment for around $5$ hours where the MCMC with HMC runs for $45,000$ iterations and the MCMC with `univariate' kernel runs for $3,000,000$ iterations.
	We run the BRCA1 experiment for around $6$ hours where the MCMC with HMC runs for $20,000$ iterations and the MCMC with `univariate' kernel runs for $600,000$ iterations.
	In the large MPXV example, the univariate kernel achieves $0.0062$ ESS/min for the dimension with the lowest effective sample size, while the HMC kernel achieves $12.55$ ESS/min, corresponding to a relative speedup of approximately $2026$-fold.
	In the small BRCA1 example, the univariate kernel achieves $2.88$ ESS/min for the dimension with the lowest effective sample size, while the HMC kernel achieves $4.64$ ESS/min, corresponding to a relative speedup of approximately $1.6$-fold. 
	We perform all benchmarking experiments utilizing $12$ CPU threads for BEAGLE calculations on a Dell Precision 5820 Tower workstation with an Intel Xeon W-2245 CPU and 128 Gb memory.
	Figure~\ref{fig:ESS} illustrates the branch-specific substitution parameter estimates binned by the ESS per minute for each dimension for the two data examples.
	The efficiency of HMC is overwhelmingly better in the MPXV example, but only slightly better compared to the univariate sampler in the BRCA1 example largely due to the considerable difference in data set size.
\begin{figure}
	\begin{center}
		\includegraphics[scale=0.6]{ESS\_v2.pdf}
	\end{center}
	\caption{
		Posterior sampling efficiency on all branch-specific substitution parameters for the BRCA1 and MPXV examples. 
		We bin parameters by their $\log_2(\mbox{ESS/min})$ values.
		The univariate and the HMC  transition kernels employed in the MCMC are color-coded.
	}
	\label{fig:ESS}
\end{figure}
%
%
%
%
%
%
%

	\section{Discussion}

	We presented a new algorithm for calculating the gradient of the phylogenetic model likelihood w.r.t.~branch-specific substitution process parameters.
	Our approach achieves linear complexity in the number of sequences by extending the post-order traversal in Felsenstein's pruning algorithm and our previous approach that calculates the gradient w.r.t.~the branch lengths \citep{Felsenstein1973, Felsenstein1981, ji2020gradients}.
	The new algorithm enables the application of gradient-based methods for optimization such as the BFGS and sampling such as HMC on learning these high-dimensional branch-specific substitution parameters.
	Interestingly, previous work achieved impressive progress using fast approximations of such gradients with bounded error \citep{magee2024random, didier2024surprising} on sampling using HMC.
	While our new algorithm provides exact analytic gradient calculations with a computational complexity of $O(\nTips m^3)$ compared to the $O(\nTips m^2)$ complexity for the approximate gradient algorithms, the efficiency trade-off of the numerical error and computational burden for the Hamiltonian dynamic integration that in-return affect HMC's sampling efficiency remains to be investigated.
	A direct and additional application of our new exact gradient algorithm is for optimization that typically involves obtaining MLEs, maximum a posterior estimates and variational analyses.

	In \cite{ji2020gradients}, we introduced a linear-time algorithm that calculates the likelihood and its gradient w.r.t.~all branch lengths through the post-order and the complementary pre-order traversal.
	In this work, we presented another linear-time algorithm that calculates the gradient w.r.t.~all branch-specific substitution parameters through the same pre-order traversal.
	In fact, the post- and pre-order partial likelihood vectors (i.e.,~$\p{i}$ and $\q{i}$) are exactly the same in both cases and therefore only need one single update.
	The difference between the two gradient calculations is in the final reduction in the numerator of Equation~\ref{Eq:singleRateCaseGradient} and Equation~\ref{Eq:GradientWithGamma} where the sandwich calculation utilizes the differential matrix and the post- and pre-order partial likelihood vectors.
	Because of this, one can calculate multiple $O(\nTips)$-dimensional gradients (for different substitution parameters and branch lengths) in one single pre-order traversal in practice.
	For derivative w.r.t.~branch length, the differential matrix (i.e.,~the middle matrix in the sandwich) is simply the instantaneous transition matrix $\mathbf{Q}_i$ where $\mathbf{Q}_i$ is independent from the branch length.
	However, when calculating the derivative w.r.t.~parameter within the generator matrix $\mathbf{Q}_i$, one needs to construct the directional derivative of the matrix exponential as in \cite{najfeld1995derivatives}.
	To match the sandwich calculation as in \cite{ji2020gradients} and make the derivative of the pre-order partial likelihood vector the same operation as left-multiplying another matrix to itself so that one re-uses the pre-order partial likelihood vectors, we need to right-multiply the inverse of the matrix exponential (i.e.,~the transition probability matrix) to the directional derivative matrix.
	This seemingly additional calculation actually results in reduced scalar exponential calculations in the construction of the middle matrix in the sandwich calculation (e.g.,~compare Equation~\ref{eq:Psi} with Equation~\ref{eq:phi}).

	Our gradient algorithm requires the transition probability matrices and their eigen decomposition being readily available after the likelihood calculation through post-order traversals (e.g.,~as in Equation~\ref{Eq:singleRateCaseGradient} and Equation~\ref{Eq:GradientWithGamma}).
	Since we compute the transition probability matrices using eigen decomposition—which has a computational cost of $O(m^3)$,
	where $m$ is the size of the state space and $m \times m$ is the dimension of the matrix—and because each branch-specific substitution parameter requires a separate decomposition, the total computational cost (including both eigen decomposition and matrix exponentiation) becomes significant, especially for large state spaces.
	This is the case for the BRCA1 data example where $m = 61$ and $\nTips = 8$ and this may explain why HMC's performance is less impressive compared to the MPXV example where $m=4$ and $\nTips = 138$.
	Fortunately, these large state space models usually lead to sparse instantaneous transition matrices (i.e.,~$\mathbf{Q}_i$ is sparse), allowing one to update the post- and pre-order partial likelihood vectors via the matrix action calculations that significantly reduce the computational cost from cubic $O(m^3)$ down to near quadratic $O(fm^2)$ depending on the sparsity $f$ of $\mathbf{Q}_i$ \citep{al2011computing, ji2016phylogenetic, sherlock2021direct}.
	However, the proposed gradient algorithm will not be readily applicable for these matrix action calculations because neither the transition probability matrices nor their eigen decompositions are available such that novel algorithms need to be developed to accommodate these promising techniques.

	A caveat of our MLE optimization comparison is that it does not include other widely used optimization criteria.
For example, the ``virtual root'' method employed by several popular phylogenetic software packages such as PhyML \citep{boussau2006efficient, guindon2010new} and RAxML \citep{stamatakis2014raxml}.
The central finite difference numerical method has a computational complexity of $O(\nTips^2)$ and over-estimates the cost for computing gradient numerically.
The virtual root method, on the other hand, achieves a lower computational complexity of $O(\nTips)$ by relocating the root adjacent to the branch being altered.
By systematically moving the root to each branch in the tree, only one partial likelihood vector needs to be recomputed at a time (effectively constructing the pre-order partial likelihood vectors).
When such vectors are available (e.g., via the virtual root traversal), our analytic gradient algorithm achieves comparable computational cost while offering higher accuracy.
Another caveat of our MLE optimization comparison is that there lacks a thorough investigation on external factors that may influence run time.
\myedit{numericVariation}{%
For instance, variations in run time may stem from differences in starting parameter values, computing hardware structures (e.g., x86 vs. ARM), operating systems, eigendecomposition methods or packages, or even compiler optimization flags (such as ``-O2'' versus ``-O3'').  %
Although randomizing the starting point may help isolate such influences, and repeating the optimization from previous end points could help ensure convergence to a local optimum (e.g.,~reinitializing the BFGS Hessian matrix approximation), such analyses are beyond the scope of the current work.%
}
\myedit{numericalError}{%
Instead, our comparison focuses on per-iteration performance, which more directly reflects differences in algorithmic efficiency (while ignoring differences in
the line-search step).
Interestingly, full-run comparisons from our limited experiments suggest greater numerical instability in the numerical gradient method, as evidenced by a notably lower maximized log-likelihood value for the MPXV example and fewer iterations for both examples.  %
This instability likely arises from the combined effects of eigen decomposition and partial likelihood vector updates throughout the tree.  %
As an exploratory investigation, we looked into the influence of varying step sizes on the numerical central difference gradient algorithm, the MLE performance of branch-specific APOBEC and the final optimized log likelihood values on simulated non-parametric bootstrap samples of the BRCA1 data example in Supplementary Material.  %
While our exploratory analyses constitute a small step into characterizing the properties of these maximum likelihood estimation procedures, we believe a thorough investigation remains an important avenue for future research.
}


	Branch-specific substitution processes have been a popular choice for modeling evolutionary heterogeneity across lineages and have led to various important studies on detecting positive selection pressures in lineages \citep{muse1994likelihood, yang1998likelihood, yang1998synonymous, yang2002codon, murrell2013fubar}.
However, most methods typically rely on a fixed tree topology and either specify {\it a priori} clusters of branches on a phylogeny to share the same substitution process or discourage using many free parameters when assuming that each branch has a specific substitution process \citep[e.g., PAML currently allows a maximum of $8$ branch types with different $dN/dS$ ratios;][]{alvarez2023beginner}.
We explore a Bayesian phylogenetic inference framework that simultaneously employs branch-specific substitution processes while searching the tree topologies without any {\it a priori} assumptions on any branches to share the same processes (which is however a special case of our model).
	The Bayesian bridge shrinkage prior that we place on the increments of branch-specific substitution parameters shows promising potential towards an ``automated'' change-point solution such that it penalizes the log difference of the substitution parameter of a branch to the parameter of the parent branch while its heavy-tail allows actual signals to pass through.
	Except for its heavier tails, the Bayesian bridge shrinkage prior works similarly to the normal priors employed in the auto-correlated branch rate models \citep{thorne1998estimating} and its counterpart in maximum likelihood methods known as a `penalty' term.
	The combination of branch-specific substitution parameters and Bayesian bridge shrinkage priors reveals clear change-points in the mutational and selection pressure dynamics in the MPXV and BRCA1 examples respectively without {\it a priori} clustering branches into discrete levels.
	While incorporating Markov-modulated modeling \citep{baele2021markov} across sites could potentially extend our approach to account for both branch- and site-specific heterogeneity, the potential identifiability issues and their resolution remain another important avenue for future research.


	
	\section{Acknowledgments}
	
	We thank Dr. Kenneth McLaughlin for thoughtful discussions.
	MAS and XJ are partially supported by NIH grants U19 R01 AI135995, AI153044 and R01 AI162611.
	XJ acknowledges support from the RCS program of Louisiana Board of Regents grant, NSF DEB1754142 and R01GM072562.
	WMD was supported by the National Institutes of Health grants CA227789,CA224381,CA287524,GM072562.
	GB acknowledges support from the Research Foundation - Flanders (``Fonds voor Wetenschappelijk Onderzoek - Vlaanderen,'' G0E1420N, G098321N), from the European Union Horizon 2023 RIA project LEAPS (grant agreement no. 101094685) and from the DURABLE EU4Health project 02/2023-01/2027, which is co-funded by the European Union (call EU4H-2021-PJ4) under Grant Agreement No. 101102733.
	PL acknowledges support by the Research Foundation -- Flanders (`Fonds voor Wetenschappelijk Onderzoek -- Vlaanderen', G066215N, G0D5117N and G0B9317N).
	We gratefully acknowledge support from NVIDIA Corporation and Advanced Micro Devices, Inc.~with the donation of parallel computing resources used for this research.

	
	
	
	\clearpage
	\bibliographystyle{natbib}
	\bibliography{branchSpecific}
\end{document}


\singlespacing

\bigskip
\medskip
\begin{center}

\noindent{\Large \bf
	Supplementary Text for:\\
	Detecting Evolutionary Change-Points with Branch-Specific Substitution Models and Shrinkage Priors
}
\bigskip

\begin{center}
	{Xiang Ji$^{\ast, 1}$, Benjamin Redelings$^{2}$, Shuo Su$^{3}$, Hongcun Bao$^{4}$, Wu-Min Deng$^{4}$, Filippo Monti$^{5}$, Samuel L. Hong$^{6}$, Guy Baele$^{6}$, Philippe Lemey$^{6}$, and Marc A. Suchard$^{\ast, 5, 7, 8}$\\\vspace{1cm}
		
		\small $^{1}$Department of Statistics, College of Liberal Arts and Sciences,\\
		Iowa State University, Ames, IA, USA\\
		\small $^{2}$Department of Mathematics, School of Science and Engineering,\\
		Tulane University, New Orleans, LA, USA\\
		\small $^{3}$Shanghai Institute of Infectious Disease and Biosecurity, School of Public Health,\\
		Fudan University, Shanghai, China\\
		\small $^{4}$Department of Biochemistry and Molecular Biology,\\
		Tulane University, New Orleans, LA, USA\\
		\small $^{5}$Department of Biostatistics, Fielding School of Public Health,\\
		University of California Los Angeles, Los Angeles, CA, USA\\
		\small $^{6}$Department of Microbiology, Immunology and Transplantation, Rega Institute,\\
		KU Leuven, Leuven, Belgium\\
		\small $^{7}$Department of Biomathematics and
		\small $^{8}$Department of Human Genetics,\\
		David Geffen School of Medicine,
		University of California Los Angeles, Los Angeles, CA, USA\\
		\small $^{*}$Correspondence: xiangji@iastate.edu, msuchard@ucla.edu
}\end{center}

\end{center}

\vspace{1in}

\clearpage

\doublespacing

\beginsupplement

\section{The derivative of matrix exponential with real and complex eigenvalues}
\myedit{ComplexGradientOne}{%
	For ease of presentation, we focus our derivation on a single infinitesimal generator matrix $\mathbf{Q} \in \mathbb{R}^{m \times m}$ where $m$ is the size of the state space.
	Our derivation is general and applies for each of the branch-specific infinitesimal rate matrix $\mathbf{Q}_i$.
	In the main text, we derive the analytical exact derivative of matrix exponential when all eigenvalues are real.
	Here we extend the derivative formula to include possible complex conjugate eigenvalues using real Schur decomposition.
}

\subsection{Real Schur decomposition}
\myedit{ComplexGradientTwo}{%
	For real matrix $\mathbf{Q} \in \mathbb{R}^{m \times m}$, there exists a real Schur decomposition $\mathbf{Q} = \mathbf{U} \mathbf{\Lambda}\mathbf{U}^{-1}$ where $\mathbf{U} \in \mathbb{R}^{m \times m}$ is a real eigenvector matrix and $\mathbf{\Lambda} \in \mathbb{R}^{m \times m}$ is a real block diagonal matrix \cite[see~e.g.,~Section~2.3 of][]{horn2012matrix}.
	The block diagonal matrix $\mathbf{\Lambda}$ consists of a $1 \times 1$ block of ${\Lambda}_{j, j} = \lambda_j$ when the eigenvalue $\lambda_j$ is real and a $2 \times 2$ block of 
	\begin{equation}
		\left(\begin{array}{cc}
			\Lambda_{j, j} & \Lambda_{j, j + 1}\\
			\Lambda_{j + 1, j} & \Lambda_{j + 1, j + 1}\\
		\end{array} \right)= \left(\begin{array}{cc}
			a_j & b_j\\
			-b_j & a_j\\
		\end{array} \right)
	\end{equation}
	for a conjugate pair of complex eigenvalues $\lambda_j = a_j + i b_j$ and $\lambda_{j + 1} = a_j - i b_j$.
}

\subsection{Matrix exponential with real and complex eigenvalues}
\label{sec:complexMatrixExponential}
\myedit{ComplexGradientThree}{%
	Using the real Schur decomposition $\mathbf{Q} = \mathbf{U} \mathbf{\Lambda}\mathbf{U}^{-1}$ above, the matrix exponential of $e^{\mathbf{Q}t}$ becomes \cite[see~e.g.,~Algorithm 10.23 -- Schur–Parlett algorithm for matrix exponential in][]{higham2008functions}
	\begin{equation}
		e^{\mathbf{Q}t} = \mathbf{U} e^{\mathbf{\Lambda}t} \mathbf{U}^{-1}
	\end{equation}
	where $e^{\mathbf{\Lambda}t}$ is a block diagonal matrix with a $1 \times 1$ block of $e^{\lambda_j t}$ when the eigenvalue $\lambda_j$ is real and a $2 \times 2$ diagonal block of  
			\begin{equation}
				{\left(\begin{array}{cc}
						e^{a_j t}\cos(b_jt) & e^{a_j t}\sin(b_jt)\\
						-e^{a_j t}\sin(b_jt) & e^{a_j t}\cos(b_jt)\\
					\end{array} \right)}
			\end{equation}
			for a conjugate pair of complex eigenvalues $\lambda_j = a_j + i b_j$ and $\lambda_{j + 1} = a_j - i b_j$.
			This is the case implemented in the BEAST X and BEAGLE software packages \citep{baele2025, ayres2019beagle}.
		}%

		\subsection{Derivative of matrix exponential with real and complex eigenvalues}
		\myedit{ComplexGradientFour}{%
			We follow the same ``spectral representations'' approach to calculate the derivative of the matrix exponential as in \cite{najfeld1995derivatives}.
			However, to include possible complex eigenvalues, we start directly with the convolution form of the derivative}
		\begin{equation}
			\label{eq:derivative_convolution}
			\begin{aligned}
				\frac{\partial \mathbf{P}}{\partial \theta} &= \frac{\partial e^{\mathbf{Q}t}}{\partial \theta}\\
				&= \int_{0}^{t} e^{(t - \tau) \mathbf{Q}} \frac{\partial \mathbf{Q}}{\partial \theta} e^{\tau \mathbf{Q}} d\tau\\
				&= \int_{0}^{t} \mathbf{U} e^{(t - \tau) \mathbf{\Lambda}} \mathbf{U}^{-1} \frac{\partial \mathbf{Q}}{\partial \theta} \mathbf{U} e^{\tau \mathbf{\Lambda}} \mathbf{U}^{-1} d\tau\\
				&= \mathbf{U} \left[ \int_{0}^{t}  e^{(t - \tau) \mathbf{\Lambda}} \bar{\mathbf{Q}} e^{\tau \mathbf{\Lambda}}  d\tau \right] \mathbf{U}^{-1}, \\
			\end{aligned}
		\end{equation}
		\myedit{ComplexGradientFive}{%
			where $ \bar{\mathbf{Q}} =  \mathbf{U}^{-1} \frac{\partial \mathbf{Q}}{\partial \theta} \mathbf{U}$ is the same as in the main text and $\bar{\mathbf{Q}}$, $\mathbf{U}$, $\mathbf{U}^{-1}$, $e^{(t - \tau) \mathbf{\Lambda}}$ and $e^{\tau \mathbf{\Lambda}}$ are all real matrices.
			Interestingly, to reuse the pre-order partial likelihood vectors in our linear-time gradient algorithm, we are not interested in the derivative matrix itself, but rather $\mathbf{P}^{-1}\frac{\partial \mathbf{P}}{\partial \theta}$ as in Equation~$9$ of the main text.
			Using the convolution of Equation~\ref{eq:derivative_convolution}, we have}
		\begin{equation}
			\begin{aligned}
				\mathbf{P}^{-1}\frac{\partial \mathbf{P}}{\partial \theta} &= \mathbf{U}e^{-\mathbf{\Lambda}t}\mathbf{U}^{-1} \cdot \mathbf{U} \left[ \int_{0}^{t}  e^{(t - \tau) \mathbf{\Lambda}} \bar{\mathbf{Q}} e^{\tau \mathbf{\Lambda}}  d\tau \right] \mathbf{U}^{-1}\\
				&= \mathbf{U} \left[ \int_{0}^{t}  e^{- \tau \mathbf{\Lambda}} \bar{\mathbf{Q}} e^{\tau \mathbf{\Lambda}}  d\tau \right] \mathbf{U}^{-1}\\
				&= \mathbf{U} \mathbf{\Xi} \mathbf{U}^{-1}, \\
			\end{aligned}
		\end{equation}
		\myedit{ComplexGradientSix}{%
			where $\mathbf{\Xi} = {\int_0^t}  e^{- \tau \mathbf{\Lambda}} \bar{\mathbf{Q}} e^{\tau \mathbf{\Lambda}}  d\tau$.
			Applying the block diagonal matrices of $e^{- \tau \mathbf{\Lambda}}$ and $e^{\tau \mathbf{\Lambda}}$ from Section~\ref{sec:complexMatrixExponential}, $\Xi_{j k}$ results in four cases: (i) $\lambda_j$ and $\lambda_k$ are both real; (ii) $\lambda_j = a_j + i b_j$ is complex and $\lambda_k$ is real; (iii) $\lambda_j$ is real and $\lambda_k = a_k + ib_k$ is complex; and (iv) $\lambda_j = a_j + i b_j$ and $\lambda_k = a_k + i b_k$ are both complex.
			Since complex eigenvalues always form $2 \times 2$ conjugate blocks, we consider them together when showing the results for cases involving complex eigenvalues.%
		}

		\paragraph{Case (i) $\lambda_j \in  \mathbb{R}$, $\lambda_k \in \mathbb{R}$.}
		\begin{equation}
			\begin{aligned}
				\Xi_{jk} &=  \int_{0}^{t}  e^{- \tau \lambda_j} \bar{Q}_{jk} e^{\tau {\lambda_k}}  d\tau\\
				&= \left\{\begin{array}{ll}
					\bar{Q}_{jk} \left[1 - e^{(\lambda_k - \lambda_j)t}\right] / (\lambda_j - \lambda_k) & \text{when } \lambda_j \neq \lambda_k\\
					\bar{Q}_{jk} t & \text{when } \lambda_j = \lambda_k\\
				\end{array} \right.\\
			\end{aligned}
		\end{equation}
		\myedit{ComplexGradientSeven}{%
		is the same as in the main text.%
}

		\paragraph{Case (ii) $\lambda_j = a_j + i b_j$, $\lambda_{j + 1} = a_j - i b_j$, $\lambda_k \in \mathbb{R}$.}
		\begin{equation}
			\begin{aligned}
				\left(\begin{array}{c}
					\Xi_{jk}\\
					\Xi_{j+1, k}\\
				\end{array}\right) &=  \int_{0}^{t} e^{-a_j \tau} \left(\begin{array}{cc}
					\cos(b_j \tau) & -\sin(b_j \tau)\\
					\sin(b_j \tau) & \cos(b_j \tau)\\
				\end{array}\right) \left(\begin{array}{c}
					\bar{Q}_{jk}\\
					\bar{Q}_{j + 1, k}\\
				\end{array}\right) e^{{\lambda_k}\tau }  d\tau\\
				&=   \left(\begin{array}{c}
					\int_{0}^{t} e^{(\lambda_k - a_j)\tau}\left[\bar{Q}_{jk} \cos(b_j\tau) - \bar{Q}_{j+1, k} \sin(b_j\tau)\right] d\tau \\
					\int_{0}^{t} e^{(\lambda_k - a_j)\tau} \left[\bar{Q}_{jk} \sin(b_j\tau) + \bar{Q}_{j + 1, k} \cos(b_j\tau)\right] d\tau\\
				\end{array}\right).
			\end{aligned}
		\end{equation}

		\paragraph{Case (iii) $\lambda_j \in \mathbb{R}$,  $\lambda_k = a_k + i b_k$, $\lambda_{k + 1} = a_{k+1} - i b_{k + 1}$.}
		\begin{equation}
			\begin{aligned}
				\left(\begin{array}{cc}
					\Xi_{jk}
					&\Xi_{j, k + 1}\\
				\end{array}\right) &=  \int_{0}^{t} e^{-\lambda_j \tau}  \left(\begin{array}{cc}
					\bar{Q}_{jk}
					&\bar{Q}_{j, k+1}\\
				\end{array}\right) \left(\begin{array}{cc}
					\cos(b_k \tau) & \sin(b_k \tau)\\
					-\sin(b_k \tau) & \cos(b_k \tau)\\
				\end{array}\right)e^{a_k \tau}  d\tau\\
				&=   \left(\begin{array}{c}
					\int_{0}^{t} e^{(a_k - \lambda_j)\tau}\left[\bar{Q}_{jk} \cos(b_k\tau) - \bar{Q}_{j, k + 1} \sin(b_k\tau)\right] d\tau \\
					\int_{0}^{t} e^{(a_k - \lambda_j)\tau} \left[\bar{Q}_{jk} \sin(b_k\tau) + \bar{Q}_{j, k + 1} \cos(b_k\tau)\right] d\tau\\
				\end{array}\right)\transpose.
			\end{aligned}
		\end{equation}

		\paragraph{Case (iv) $\lambda_j = a_j + i b_j$, $\lambda_{j + 1} = a_j - i b_j$,  $\lambda_k = a_k + i b_k$, $\lambda_{k + 1} = a_k - i b_k$.}
		\begin{equation}
			\begin{aligned}
				&\left(\begin{array}{cc}
					\Xi_{jk}		&\Xi_{j, k + 1}\\
					\Xi_{j + 1, k}		&\Xi_{j + 1, k + 1}\\
				\end{array}\right)\\
				=&  \int_{0}^{t} e^{-a_j \tau} \left(\begin{array}{cc}
					\cos(b_j \tau) & -\sin(b_j \tau)\\
					\sin(b_j \tau) & \cos(b_j \tau)\\
				\end{array}\right)  \left(\begin{array}{cc}
					\bar{Q}_{jk}	&\bar{Q}_{j, k+1}\\
					\bar{Q}_{j + 1, k}	&\bar{Q}_{j + 1, k+1}\\
				\end{array}\right) \left(\begin{array}{cc}
					\cos(b_k \tau) & \sin(b_k \tau)\\
					-\sin(b_k \tau) & \cos(b_k \tau)\\
				\end{array}\right)e^{a_k \tau}  d\tau,\\
			\end{aligned}
		\end{equation}
		such that
		\begin{equation}
			\begin{aligned}
				\Xi_{jk} =& \int_{0}^{t} e^{(a_k-a_j) \tau}\left[\bar{Q}_{jk}\cos(b_j\tau)\cos(b_k \tau) - \bar{Q}_{j + 1, k}\sin(b_j\tau)\cos(b_k \tau) \right.\\
				&\left.- \bar{Q}_{j,k + 1}\cos(b_j\tau)\sin(b_k \tau) + \bar{Q}_{j+1, k+1}\sin(b_j\tau)\sin(b_k \tau) \right]d\tau\\
				=&\frac{1}{2} \int_{0}^{t} e^{(a_k-a_j) \tau}\left\{(\bar{Q}_{jk} - \bar{Q}_{j+1, k+1})\cos\left[(b_j + b_k)\tau\right] + (\bar{Q}_{jk} + \bar{Q}_{j+1, k+1})\cos\left[(b_j - b_k)\tau\right]  \right.\\
				&\left. -(\bar{Q}_{j+1, k} + \bar{Q}_{j, k+1})\sin\left[(b_j + b_k)\tau\right] + (\bar{Q}_{j, k+1} - \bar{Q}_{j+1, k})\sin\left[(b_j - b_k)\tau\right]   \right\}d\tau,\\
			\end{aligned}
		\end{equation}
		\begin{equation}
			\begin{aligned}
				\Xi_{j, k+1} =& \int_{0}^{t} e^{(a_k-a_j) \tau}\left[\bar{Q}_{jk}\cos(b_j\tau)\sin(b_k \tau) - \bar{Q}_{j + 1, k}\sin(b_j\tau)\sin(b_k \tau) \right.\\
				&\left.+ \bar{Q}_{j,k + 1}\cos(b_j\tau)\cos(b_k \tau) - \bar{Q}_{j+1, k+1}\sin(b_j\tau)\cos(b_k \tau) \right]d\tau\\
				=&\frac{1}{2} \int_{0}^{t} e^{(a_k-a_j) \tau}\left\{(\bar{Q}_{j, k+1} + \bar{Q}_{j+1, k})\cos\left[(b_j + b_k)\tau\right] + (\bar{Q}_{j, k+1} - \bar{Q}_{j+1, k})\cos\left[(b_j - b_k)\tau\right]  \right.\\
				&\left. +(\bar{Q}_{j, k} - \bar{Q}_{j+1, k +1})\sin\left[(b_j + b_k)\tau\right] - (\bar{Q}_{j, k} + \bar{Q}_{j+1, k+1})\sin\left[(b_j - b_k)\tau\right]   \right\}d\tau,\\
			\end{aligned}
		\end{equation}
		\begin{equation}
			\begin{aligned}
				\Xi_{j+1, k} =& \int_{0}^{t} e^{(a_k-a_j) \tau}\left[\bar{Q}_{jk}\sin(b_j\tau)\cos(b_k \tau) + \bar{Q}_{j + 1, k}\cos(b_j\tau)\cos(b_k \tau) \right.\\
				&\left.- \bar{Q}_{j,k + 1}\sin(b_j\tau)\sin(b_k \tau) - \bar{Q}_{j+1, k+1}\cos(b_j\tau)\sin(b_k \tau) \right]d\tau\\
				=&\frac{1}{2} \int_{0}^{t} e^{(a_k-a_j) \tau}\left\{(\bar{Q}_{j, k+1} + \bar{Q}_{j+1, k})\cos\left[(b_j + b_k)\tau\right] + (\bar{Q}_{j +1, k} - \bar{Q}_{j, k + 1})\cos\left[(b_j - b_k)\tau\right]  \right.\\
				&\left. +(\bar{Q}_{j, k} - \bar{Q}_{j+1, k +1})\sin\left[(b_j + b_k)\tau\right] + (\bar{Q}_{j, k} + \bar{Q}_{j+1, k+1})\sin\left[(b_j - b_k)\tau\right]   \right\}d\tau,\\
			\end{aligned}
		\end{equation}
		\begin{equation}
			\begin{aligned}
				\Xi_{j+1, k+1} =& \int_{0}^{t} e^{(a_k-a_j) \tau}\left[\bar{Q}_{jk}\sin(b_j\tau)\sin(b_k \tau) + \bar{Q}_{j + 1, k}\cos(b_j\tau)\sin(b_k \tau) \right.\\
				&\left.+ \bar{Q}_{j,k + 1}\sin(b_j\tau)\cos(b_k \tau) + \bar{Q}_{j+1, k+1}\cos(b_j\tau)\cos(b_k \tau) \right]d\tau\\
				=&\frac{1}{2} \int_{0}^{t} e^{(a_k-a_j) \tau}\left\{(\bar{Q}_{j+1, k+1} - \bar{Q}_{jk})\cos\left[(b_j + b_k)\tau\right] + (\bar{Q}_{j+1, k+1} + \bar{Q}_{jk})\cos\left[(b_j - b_k)\tau\right]  \right.\\
				&\left. +(\bar{Q}_{j+1, k} + \bar{Q}_{j, k+1})\sin\left[(b_j + b_k)\tau\right] + (\bar{Q}_{j, k+1} - \bar{Q}_{j+1, k})\sin\left[(b_j - b_k)\tau\right]   \right\}d\tau.\\
			\end{aligned}
		\end{equation}
		We omit writing out each individual integral because they all take one of two forms
		\begin{equation}
			\begin{aligned}
				\int_{0}^{t} e^{x \tau}\cos(y\tau)d\tau &= \frac{ye^{xt}\sin(yt) + xe^{xt}\cos(yt) - x}{x^2+y^2}\\
				\int_{0}^{t} e^{x \tau}\sin(y\tau)d\tau &= \frac{xe^{xt}\sin(yt) - ye^{xt}\cos(yt) + y}{x^2+y^2}\\
			\end{aligned}
		\end{equation}
		\myedit{ComplexGradientEight}{%
			where $x, y \in \mathbb{R}$ are two constants and $y \neq 0$.}

\section{Simulation study}

\myedit{SimulationStudy}{%
We refer to \cite{fisher2023shrinkage} for a simulation study of the shrinkage prior performance.
We investigate the performance of the MLE procedure that depends on the combination of (1) accuracy of the gradient calculations and (2) the statistical model (branch-specific substitution process and the associated penalty terms).
To further separate these sources of influence, we conduct two separate investigations using the MPXV example.

We first explore the influence of the relative step size ratio ($r$) in the calculation of the central difference numerical gradient ($\boldsymbol{g}_{numeric} \in \mathbb{R}^n$).
The central difference numeric gradient algorithm calculates the $i$th derivative by
\begin{equation}
\left\{\boldsymbol{g}_{numeric}\right\}_i = \frac{f(\boldsymbol{x} + \Delta_x \boldsymbol{e}_i) - f(\boldsymbol{x} - \Delta_x \boldsymbol{e}_i)}{{2\Delta_x}}
\end{equation}
where $f(\boldsymbol{x})$ is the log likelihood function, $\boldsymbol{x} \in \mathbb{R}^n$ is the parameter of interest, $\boldsymbol{e}_i$ is a unit vector and $\Delta_x = r \cdot \max\{1, |x_i|\}$ is the step size for dimension $i$.
By the limit definition of a derivative (i.e.,~$\frac{\partial f(\boldsymbol{x})}{\partial x_i} = \lim\limits_{\Delta_x \to 0} \frac{f(\boldsymbol{x} + \Delta_x \boldsymbol{e}_i) - f(\boldsymbol{x} - \Delta_x \boldsymbol{e}_i)}{2\Delta_x}$), one would expect the numerical gradient to be more precise (e.g.,~closer to the analytical gradient) when the step size approaches zero.}
\renewcommand{\figurename}{Supplementary Figure}
\begin{figure}
	\begin{center}
		\includegraphics[scale=0.6]{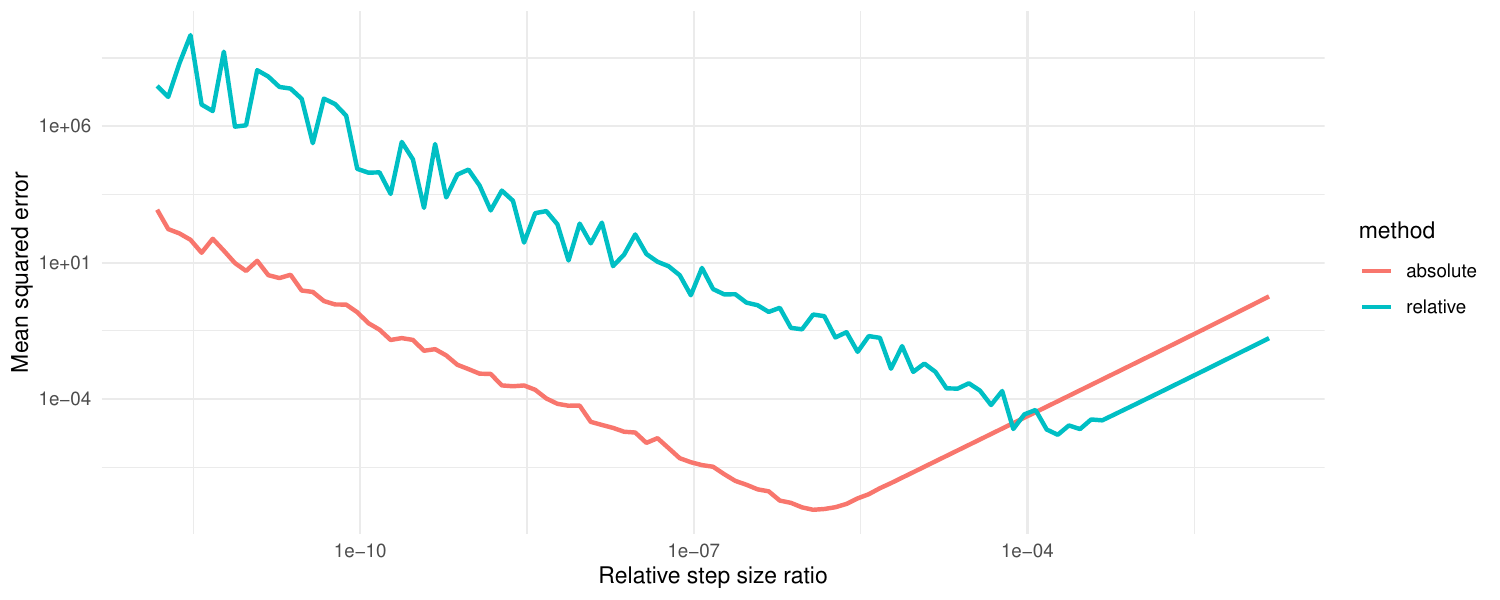}\vspace{-0.6cm}
	\end{center}
	\caption{
		Mean squared error of the central difference numerical gradient compared to the analytical gradient over step sizes.
	}
	\label{fig:mse}
\end{figure}
\myedit{SimulationStudyTwo}{%
However, this is not the case as illustrated in Supplementary Figure~\ref{fig:mse} where we compared the numerical gradient from varying the step size ratio $r$ to the analytical gradient ($\boldsymbol{g}_{analytic} \in \mathbb{R}^n$) by mean squared errors in}
\begin{enumerate}
	\item absolute values such as $\frac{1}{n}(\boldsymbol{g}_{numeric} - \boldsymbol{g}_{analytic})\transpose(\boldsymbol{g}_{numeric} - \boldsymbol{g}_{analytic})$,
	\item relative ratios such as  $\frac{1}{n}(\boldsymbol{g}_{numeric} / \boldsymbol{g}_{analytic} - 1)\transpose(\boldsymbol{g}_{numeric} / \boldsymbol{g}_{analytic} - 1)$.
\end{enumerate}
\myedit{SimulationStudyThree}{%
It is interesting that floating point error starts to affect the accuracy of the numerical gradient when the step sizes become too small.
Therefore, we primarily use {\it SQRT\_SQRT\_EPSILON} which is $1.220703125\times 10^{-4}$ (the 4th root of machine accuracy {\it EPSILON}, $2.220446049250313 \times 10^{-16}$) as a reasonable step size ratio for calculating the numerical gradient w.r.t.~substitution parameters.

We proceeded to perform an exploratory simulation study.
We used the MPXV example (chosen for its larger number of tips) with the same starting parameter values as in the Main Text except for the branch-specific APOBEC rates. %
Supplementary Figure~\ref{fig:simulation} illustrates the MLEs of log APOBEC rates for four sets of $100$ simulated datasets.
For each of the $100$ simulated datasets, we set the background APOBEC rate to $e^{0.1}$ on the tree with an accelerated rate of (A) $e^{0.4}$, (B) $e^{0.6}$, (C) $e^{0.8}$ and (D) $e^{1.0}$ after the ``star'' branch indicated in Figure~2 in the manuscript.
Each simulated alignment has the same size as the original dataset.}
\begin{figure}
	\begin{center}
		\includegraphics[scale=0.3]{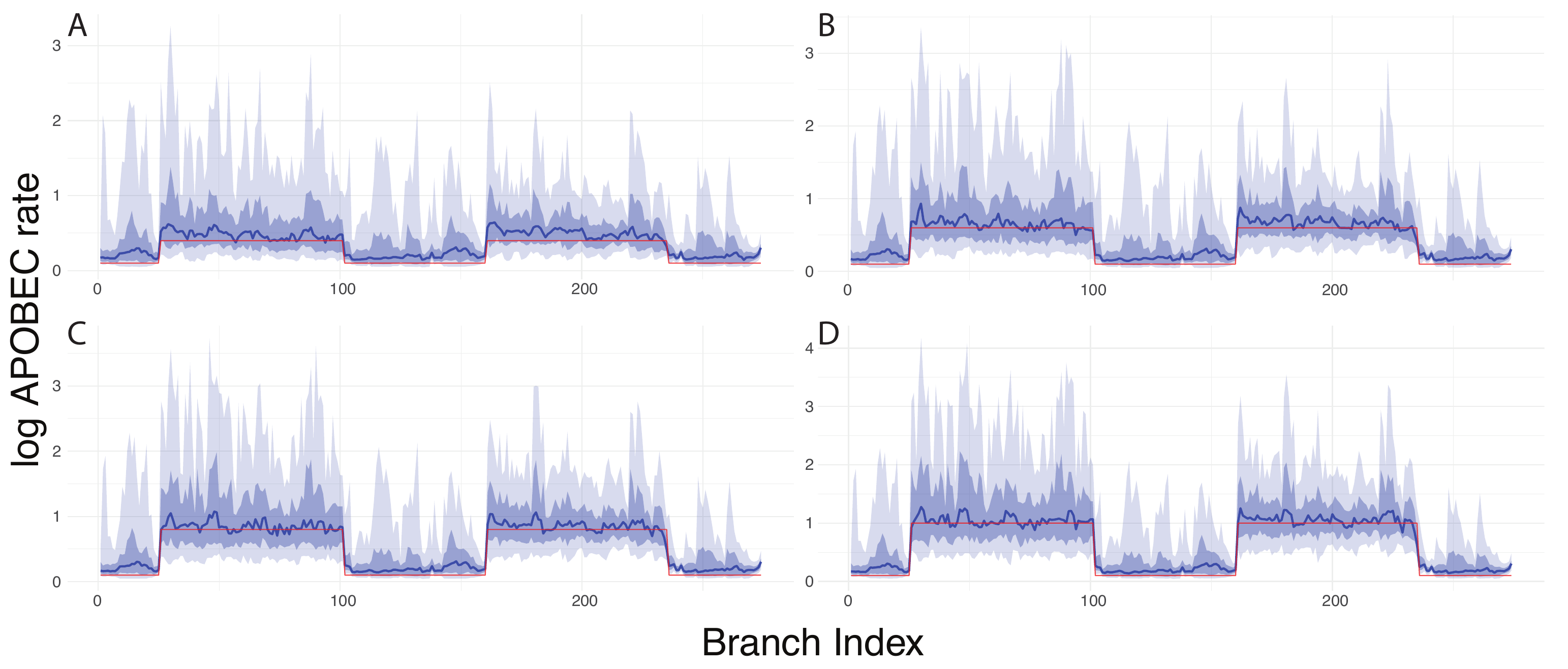}\vspace{-0.6cm}
	\end{center}
	\caption{
		Shaded quantile plots of MLEs of log APOBEC rates over branches of $4$ sets of $100$ simulated datasets.  Light blue shade indicates $5\%$ and $95\%$ quantiles.  Darker blue shade indicates $25\%$ and $75\%$ quantiles.  Solid blue lines indicate the median estimate.  Red lines indicate the true values used to simulate the datasets that are (A) $0.4$, (B) $0.6$, (C) $0.8$ and (D) $1.0$.
	}
	\label{fig:simulation}
\end{figure}
%
\myedit{SimulationStudyFour}{%
The MLEs capture the change-point locations well, but the estimates are slightly biased (as expected when using a penalty term).
This limited experiment indicates reasonable performance of the maximum-likelihood estimator; however, a thorough investigation relaxing and estimating all parameters (including the tree topology) is important for a better insight of its usage in different scenarios (e.g.,~size of trees, substitution processes, variations of branch-specific parameters, etc.).

Finally, we performed an experiment to investigate if the analytical gradient consistently results in higher maximized log likelihoods with the same optimization algorithm compared to a numerical gradient.
We created $100$ non-parametric bootstrap datasets by sampling with replacement from the original columns of the BRCA1 sequence alignment.
We then estimated MLEs of branch-specific omega parameters on these bootstrap datasets while fixing all other model parameters (including tree topology) at the same values as in \emph{Section~4.1 Optimization}.
We compare the maximized log likelihood values obtained via analytical and numeric gradients of the $100$ bootstrap datasets by their differences where a positive difference corresponds to higher maximized log likelihood value obtained using the analytical gradient.
The $100$ log likelihood differences range from a minimum of $-0.087$, first quantile of $-0.05$, median of $-4.9\times 10^{-6}$, third quantile of $0.05$ to a maximum of $5.64$ with a mean of $0.04$.
While looking at the absolute value of these log likelihood differences, they range from a minimum of $3.3\times 10^{-8}$, first quantile of $1.2\times 10^{-6}$, median of $1.0 \times 10^{-4}$, third quantile of $0.05$ to a maximum of $5.64$ with a mean of $0.08$.
Results from this experiment indicate comparable performance of the analytical and numerical gradient algorithms for optimization.}

\section{Comparison of posterior estimates}

We summarize the performance of the HMC and the univariate samplers for approximating the posterior distributions in Supplementary Figures~\ref{fig:BRCA1_violin} and \ref{fig:apobe_violin}.
We compare the distribution of the log joint density (proportional to the normalized posterior distribution) of MCMC samples for both examples via violin and box plots.
We compare the estimated marginal posterior distribution of branch-specific omega parameters for the BRCA1 example via violin and box plots in Supplementary Figure~\ref{fig:BRCA1_violin}.
We compare the estimated posterior median (solid point) and $95\%$ highest posterior density intervals (error bars) of branch-specific log APOBEC rates and their transformed increments for the MPXV example in Supplementary Figure~\ref{fig:apobe_violin}.

The distributions of log joint density using the two samplers (HMC and univariate) match well for both examples.
The estimated marginal posterior distributions for the branch-specific omega parameters match well for most of the branches for the BRCA1 example.
The estimated marginal posterior distributions for the branch-specific APOBEC rates match reasonably well for the branches with small estimated APOBEC rate values corresponding to branches before the ``star'' branch in Figure~$2$ in the main text, but illustrate a slight mismatch for branches with elevated APOBEC rates.
This mismatch is caused by one single mismatch in the increment estimates (as shown in the right panel of Supplementary Figure~\ref{fig:apobe_violin}) where the univariate sampler suffers from poor mixing  for learning this increment dimension resulting in a low ESS value of $6$ (copmaring to HMC sampler's ESS of $4051$).
We include BEAST X log files in both the GitHub repository and Dryad storage.

\begin{figure}
	\begin{center}
		\includegraphics[scale=0.65]{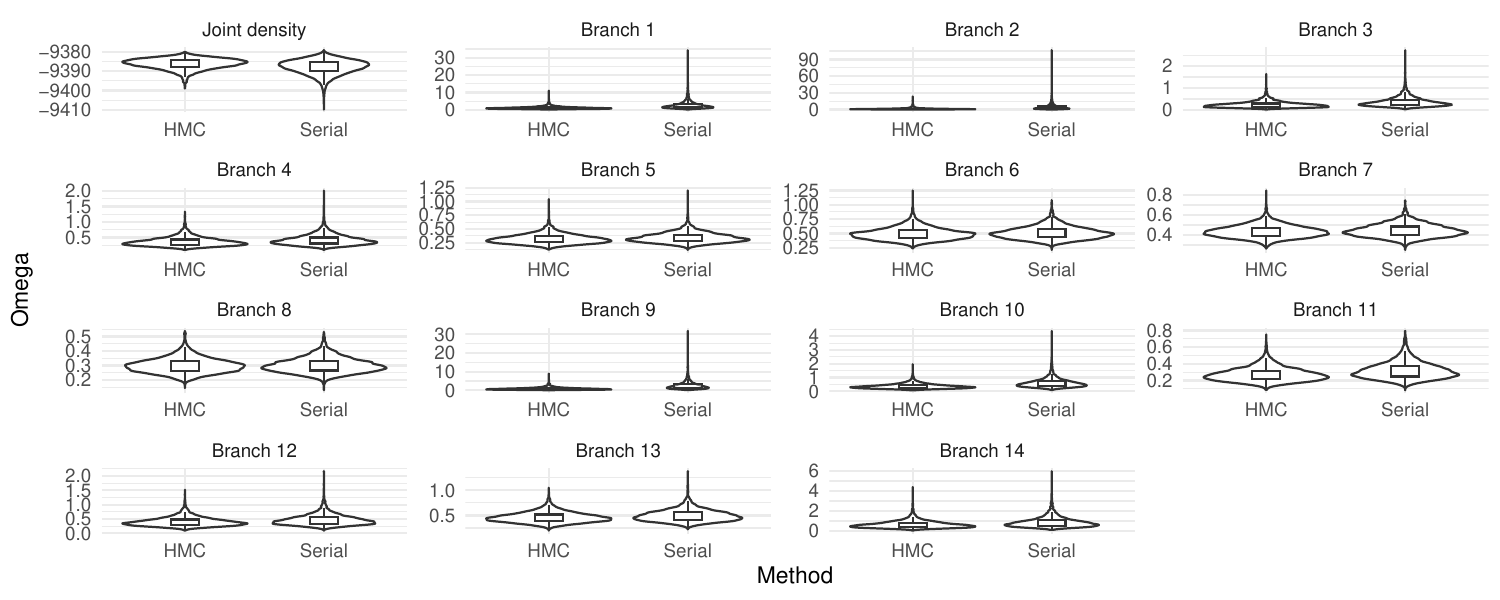}\vspace{-0.6cm}
	\end{center}
	\caption{
		Comparison of sampled log joint density distribution and posterior estimates between the HMC and the univariate samplers for the BRCA1 example.
	}
	\label{fig:BRCA1_violin}
\end{figure}
\begin{figure}
	\begin{center}
		\includegraphics[scale=0.65]{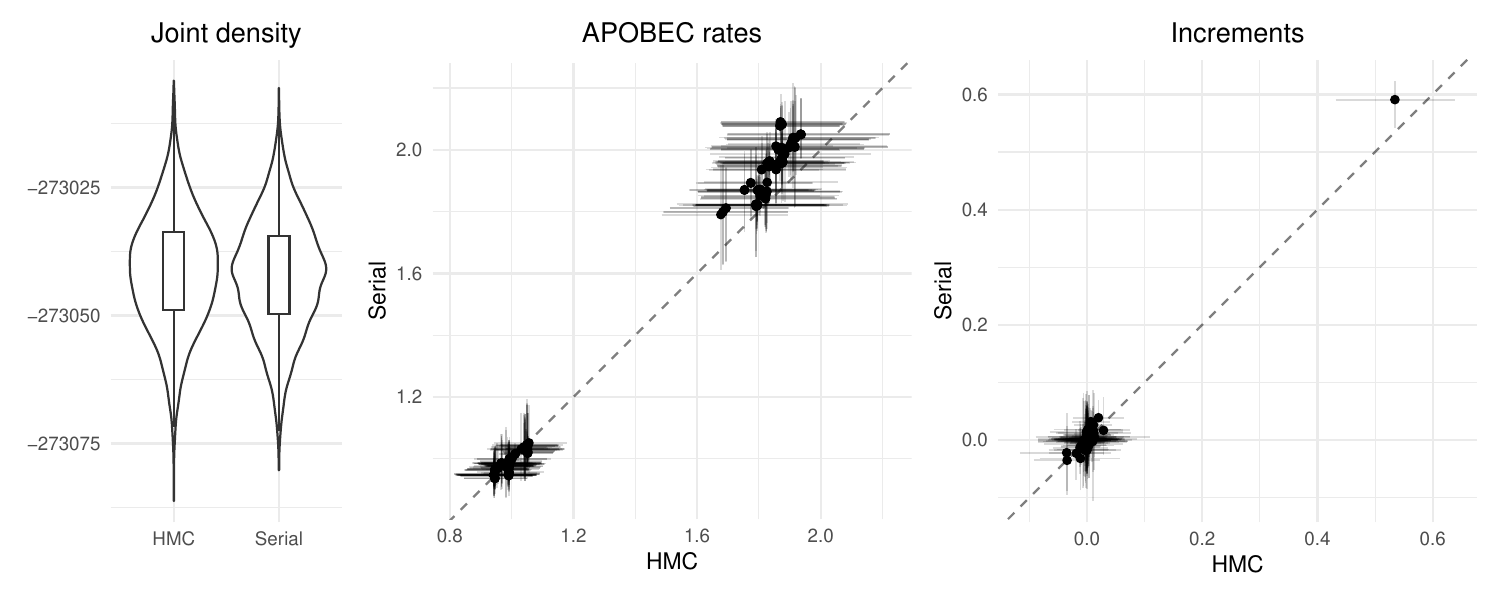}\vspace{-0.6cm}
	\end{center}
	\caption{
		Comparison of (left) sampled log joint density distribution; (middle) posterior estimates of log APOBEC rates; and (right) posterior estimation of Increments of the log APOBEC rates for the MPXV example.
		For the middle and right plots, each point represents a model parameter (log APOBEC rate or increment of log APOBEC rate) indicating posterior medians with error bars indicating $95\%$ highest posterior density intervals from the MCMC samples using the HMC sampler (X axis) and the univariate sampler (Y axis).
	}
	\label{fig:apobe_violin}
\end{figure}

\singlespacing 
\clearpage 
	\bibliographystyle{natbib}
\bibliography{branchSpecific}